\documentclass[preprint,11pt]{aastex}
\pdfoutput=1   

\usepackage{color}
\usepackage{natbib}
\usepackage{amssymb,amsmath}
\usepackage[pdftex,                
bookmarks=true,                    
bookmarksnumbered=true,            
colorlinks=true,            
citecolor=blue,         
linkcolor=blue,     
menucolor=blue,         
urlcolor=cyan,       
linkbordercolor={0 0 1},           
pdfborder= {0 0 1},
frenchlinks=True]{hyperref}
\usepackage[left=0.7in,right=1in,bottom=1in,top=1in]{geometry}

\providecommand{\eprint}[1]{\href{http://arxiv.org/abs/#1}{#1}}

\providecommand{\adsurl}[1]{\href{#1}{ADS}}
\newcommand{\gjb}{GJ~1214b}
\newcommand{\gj}{GJ~1214}
\newcommand{\rearth}{\ensuremath{R_{\earth}}}
\newcommand{\rsun}{\ensuremath{R_{\odot}}}
\newcommand{\bc}{\begin{center}}
\newcommand{\ec}{\end{center}}
\newcommand{\fig}[5]{
        \begin{figure}[!bt]
        \bc
        \includegraphics[#3]{#2}
        \ec
        \renewcommand{\baselinestretch}{1}
        \vspace*{-.3in}
        \caption[#4]{#5}
        \label{fig:#1}
        \end{figure}}

\newcommand{\figtwocol}[5]{
        \begin{figure*}[!bt]
        \bc
        \includegraphics[#3]{#2}
        \ec
        \renewcommand{\baselinestretch}{1}
        \vspace*{-.3in}
        \caption[#4]{\footnotesize #5}
        \label{fig:#1}
        \end{figure*}}

\def\aap{{A\&A}}		
\def\apj{{ApJ}}			
\def\apjl{{ApJ}}		
\def\pasp{{PASP}}		

\def\mnras{{MNRAS}}
\def\nat{{Nature}}

\def\methane{\ensuremath{\textrm{CH}_4}}
\def\coo{\ensuremath{\textrm{CO}_2}}
\def\water{\ensuremath{\textrm{H}_2\textrm{O}}}
\def\kms{\ensuremath{\textrm{km~s}^{-1}}}

\long\def\symbolfootnote[#1]#2{\begingroup%
  \def\thefootnote{\fnsymbol{footnote}}\footnote[#1]{#2}\endgroup} 

\newenvironment{my_itemize}{
\begin{itemize}
  \setlength{\itemsep}{1pt}
  \setlength{\parskip}{0pt}
  \setlength{\parsep}{0pt}}{\end{itemize}
}

\shortauthors{Crossfield et al.~2011} \shorttitle{ High-resolution spectroscopy of  \gjb }
\begin{document}


\title{High Resolution, Differential, Near-infrared Transmission Spectroscopy of \gjb
}

\slugcomment{Accepted to ApJ, 2011 May 20.}

\author{
I. J. M. Crossfield\altaffilmark{1},
Travis Barman\altaffilmark{2},
Brad M. S. Hansen\altaffilmark{1}
}

\altaffiltext{1}{Department of Physics \& Astronomy, University of California Los Angeles, Los Angeles, CA 90095, USA, ianc@astro.ucla.edu}
\altaffiltext{2}{Lowell Observatory, 1400 West Mars Hill Road, Flagstaff, AZ 86001, USA}

\begin{abstract}
  The nearby star \gj\ hosts a planet intermediate in radius and mass
  between Earth and Neptune, resulting in some uncertainty as to its
  nature.  We have observed this planet, \gjb, during transit with the
  high-resolution, near-infrared NIRSPEC spectrograph on the Keck II
  telescope, in order to characterize the planet's atmosphere. By
  cross-correlating the spectral changes through transit with a suite
  of theoretical atmosphere models, we search for variations
  associated with absorption in the planet atmosphere.  Our
  observations are sufficient to rule out tested model atmospheres
  with wavelength-dependent transit depth variations $\gtrsim 5 \times
  10^{-4}$ over the wavelength range $2.1-2.4$~\micron.  Our
  sensitivity is limited by variable slit loss and telluric
  transmission effects.

  We find no positive signatures but successfully rule out a number of
  plausible atmospheric models, including the default assumption of a
  gaseous, H-dominated atmosphere in chemical equilibrium. Such an
  atmosphere can be made consistent if the absorption due to methane
  is reduced. Clouds can also render such an atmosphere consistent
  with our observations, but only if they lie higher in the atmosphere
  than indicated by recent optical and infrared measurements.

  When taken in concert with other observational constraints, our
  results support a model in which the atmosphere of \gjb\ contains
  significant H and He, but where \methane\ is depleted. If this
  depletion is the result of photochemical processes, it may also
  produce a haze that suppresses spectral features in the optical.
\end{abstract}


\section{Introduction}

\subsection{Characterizing Extrasolar Atmospheres}
Astronomers are poised to soon discover the first rocky, Earth-like
exoplanets orbiting in the habitable zone -- the region where water
could be expected to be liquid on the planet's surface.  Already,
ground-based exoplanet surveys are finding super Earths (planets
several times Earth's mass and radius) such as \gjb\ with temperatures
only a few hundred degrees hotter than Earth \citep{charbonneau:2009}
-- much more hospitable than the more easily detected giant, highly
irradiated Hot Jupiters.  In addition, the {\em Kepler} spacecraft is
currently searching for the subtle, periodic dip in a star's
brightness that will indicate an Earth-sized planet in an Earth-like
orbit has passed in front of its Sun-like stellar host, and indeed a
preliminary list of 54 habitable planet candidates has already been
released \citep{borucki:2011}.

Transit and radial velocity measurements can constrain the radius and
mass of a planet, but significant degeneracies remain in the
determination of interior and atmospheric composition.  From
theoretical models of planetary interiors we know that any
Jupiter-sized object is composed primarily of H$_2$ and He.  A
super Earth is more complicated and thus more interesting: it could be
a small, rocky core with an extended H$_2$/He envelope (a
mini-Neptune), a larger, icy core with a denser molecular atmosphere
(a scaled-up Ganymede), or something in between \citep{rogers:2010}.
To determine the true makeup of such a planet we need precise methods
to characterize the atmospheres of these new worlds.

All else being equal, planets which transit small, low-mass stars are
the most favorable targets for atmospheric characterization because
the planet/star size ratio $R_P/R_*$ is especially high.  When the
host star occults the planet one can measure the planet's intrinsic
emission spectrum; when the planet transits its star, the transit
depth will vary with wavelength in a way that reflects absorption in
the atmosphere at the limb of the planet.  This quantity,
$(R_P(\lambda)/R_*)^2$, is often called the transmission spectrum of
the planet.  The wavelength-dependent atmospheric signature during
transit is a variable function of wavelength, but in general is
proportional to the relative area of the atmospheric annulus probed,
i.e. $H R_P(\lambda) / R_*^2$, where $H$ is the standard atmospheric
scale height \citep{miller-ricci:2009}. In the most favorable cases
the transmission spectrum's features vary by a factor of (roughly)
$10^{-3}$ relative to the flux from the planet's host star --
challenging to detect, but much more tractable than the expected
planet/star flux contrast of $\lesssim 10^{-4}$ from a relatively
temperate planet.

While numerous transit observations have been made from space, it is
only in the last few years that ground-based infrared characterization
has met with success.  Ground-based infrared photometry is becoming
almost commonplace \cite[e.g.,][]{rogers:2009, croll:2011a}, but
ground-based spectroscopy remains challenging.  \cite{swain:2010}
reported the detection of the K and L band emission spectrum of
HD~189733b, but these results are contested \citep{madhusudhan:2009,
  mandell:2011}. On a more optimistic note, the recently reported
detection of CO on HD~209458b \citep{snellen:2010} demonstrates the
power of template cross-correlation techniques for ground-based
spectroscopy. We will adopt a similar philosophy in this paper.

Recent observations of the low mass planet \gjb\ represent some of the
most precise ground-based exoplanet spectroscopy to date
\citep{bean:2010}, and show a flat transmission spectrum from
$0.75-1$~\micron .  Our study aims to complement these results in the
near-infrared.  In this paper we describe our observations of \gjb\ in
an attempt to detect the differential transmission spectrum of the
planet's atmosphere.  We describe the \gj\ system and the suite of
atmospheric models we test in Section~\ref{sec:obs}.  We discuss our
observations and data reduction in Section~\ref{sec:obsanal}, and
describe our methods for measuring relative spectrophotometry and
performing model template cross-correlations in
Section~\ref{sec:detection}.  We finish with a discussion of our
results and future work in Section~\ref{sec:results} and conclude in
Section~\ref{sec:conclusion}.

\section{Observations and Atmospheric Models }
\label{sec:obs}

\subsection{The  \gj\ System}
\gjb\ is the first planet discovered by the MEarth project
\citep{irwin:2009}, a transit survey targeting the nearest red dwarfs.
The planet orbits a 3000~K, M4.0-4.5 dwarf with high metallicity
\citep[+0.3 to +0.4;][]{schlaufman:2010,rojas-ayala:2010}.  The star
displays a periodic, 1\%\ variability in the red optical and evidence
of spots is seen during planetary transits, but the star appears to be
only weakly active and spot features should affect near-infrared
transmission spectroscopy only at the $<0.0003$ level
\citep{berta:2011} -- beneath the sensitivity we achieve in this
work. Because the star is so cool, the system is most amenable to
characterization at infrared wavelengths.  With a semimajor axis of
0.014~AU, the planet's (albedo-dependent) equlibrium temperature is
400-550~K \citep{charbonneau:2009}.

\gjb 's size and mass \citep[2.65 and 6.45 times that of the Earth,
respectively;][]{carter:2011} give it a density intermediate to that
of our solar system's inner (rocky) and outer (gas-dominated) planets.
\gjb\ occupies an intriguing location in the planetary mass-radius
diagram, inasmuch as these two bulk characteristics allow for
substantial degeneracies in models of the planet's interior
composition.  Nonetheless, models generally suggest that the planet
has a substantial gas component, making it potentially amenable to
atmospheric characterization \citep{rogers:2010, nettelmann:2010}.

Though the present constraints on \gjb 's bulk composition preclude
any unique predictions of the planet's atmospheric structure and
composition, \cite{miller-ricci:2010} predicted that the planet would
show substantial ($\le 10^{-3}$) variations in transit depth with wavelength
if it hosts a cloudless, H$_2$/He-dominated
atmosphere.  They also predicted that the atmospheric signature would
be an order of magnitude lower -- essentially undetectable in the
near-infrared -- for a denser atmosphere dominated by heavier
molecular species.  Any clouds  in \gjb 's atmosphere would
further mask the presence of any spectral signals.

The planet to star size ratio $R_P/R_*$ of the \gj\ system has been
measured photometrically in the optical
\citep{charbonneau:2009,carter:2011,berta:2011} and mid-infrared
\citep{desert:2011} and with optical spectroscopy
\citep{bean:2010}. These measurements all agree to within
$\lesssim$2\%, and have been interpreted as evidence for a clear
atmosphere dominated by heavy molecular species
\citep{desert:2011}. An atmosphere obscured by clouds or haze would
also be consistent with these observations, but a haze would need to
be composed of rather large (several microns) particles to flatten the
spectrum all the way into the mid-infrared \citep{desert:2011}.  On
the other hand, recent measurements in the near infrared
\citep{croll:2011b} argue for a larger radius at Ks-band, indicative
of a H-dominated atmosphere with large scale height. One way to make
the observations consistent with one another is to postulate a
methane-depleted atmosphere with a haze of small particles at
$R_P/R_* \approx 0.1165$ to smooth out features in the optical
\citep{desert:2011,croll:2011b}.

\subsection{Atmospheric Models of \gjb}
\label{sec:models} We test our observations (described in
Section~\ref{sec:obsanal}) by comparing them to high-resolution model
transmission spectra.  Thus our detection method is only as good as
the molecular line lists we use for our models: significant revision
of the near-infrared opacities would require a recalculation of the
constraints we ultimately present.

To explore a variety of
atmospheric conditions in \gjb\ we generated several irradiated
planetary models using the PHOENIX code, assuming traditional
H$_2$/He-rich atmospheres \citep{barman:2005}.  The baseline models
were taken to be cloud free with either solar \citep{asplund:2005},
10$\times$ solar, or 30$\times$ solar abundances (i.e., the abundance
of all elements other than H and He enhanced by the specified factor),
and we used the bulk planet and star parameters from
\cite{charbonneau:2009}.

We generated the following atmospheric models:

\begin{my_itemize}
\item Solar composition and equilibrium chemistry. 

\item 10$\times$ solar abundance and equilibrium chemistry. This model
  has a slightly smaller radius, and is hotter below 0.1~mbar,
  than the solar model.

\item 30$\times$ solar abundance and equilibrium chemistry. This model
  has a slightly higher temperature and smaller radius than the solar
  and 10$\times$ solar models.

\item No methane. Solar composition, but with the concentration of
  \methane\ set to zero.  We used this model to interpolate to various
  levels of atmospheric \methane .  This allows us to test the
  hypothesis that \methane\ abundances may be substantially reduced in
  cooler atmospheres at the pressures probed by transmission
  spectroscopy \citep{zahnle:2009,desert:2011}.

\item Low carbon abundance.  This model assumes chemical equilibrium
  but with the carbon abundance set to $10^{-8}$ relative to solar;
  this factor is roughly comparable to the ratio of the lowest and
  highest carbon-containing compounds found in our solar abundance
  model.  This model is similar to the methane-depleted model just
  described, but features from residual \methane\ are still visible
  from 2.2-2.35~\micron .


\end{my_itemize}
We plot the atmospheric temperature profiles and the abundance of
several molecular species as a function of pressure for several of
these models in Figure~\ref{fig:atmo}.

\fig{atmo}{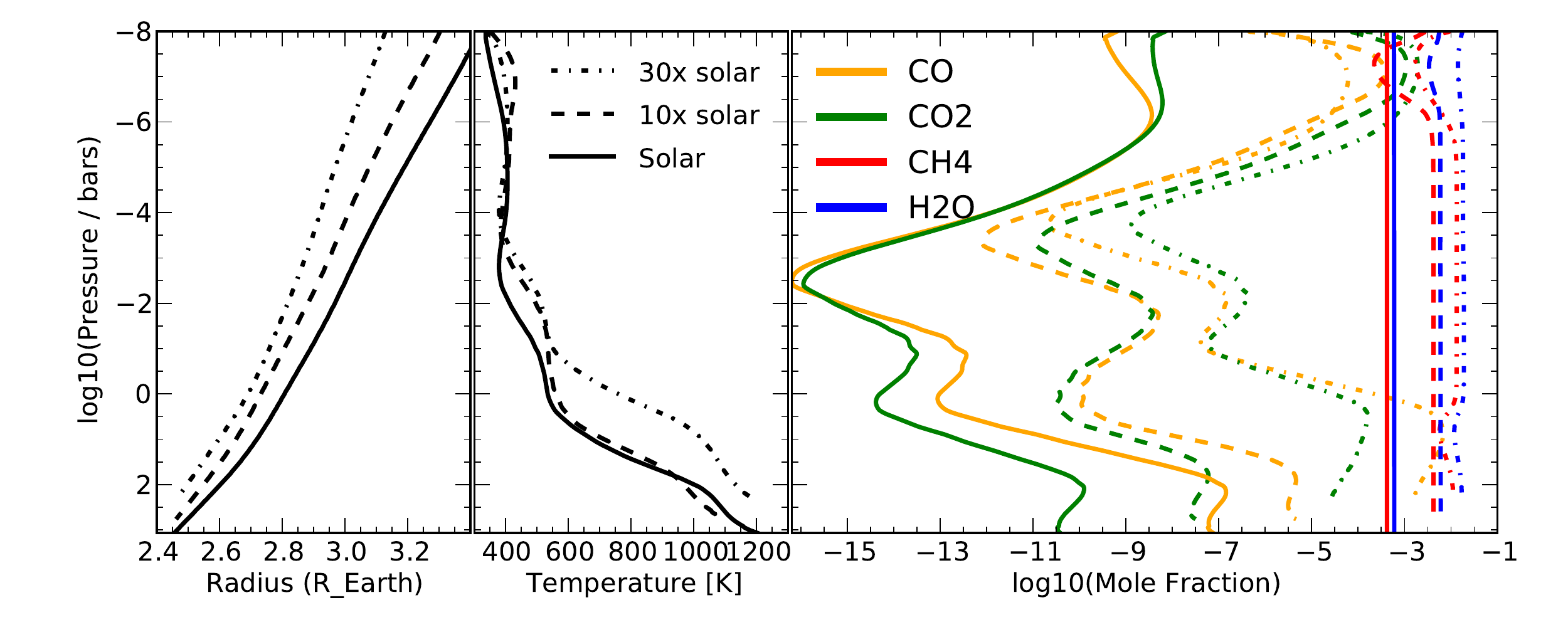}{width=7in}{}{Atmospheric parameters for our
  solar (dot-dashed), 10$\times$ solar (dashed), and 30$\times$ solar
  (solid) abundance models.  From left to right, the panels show as a
  function of pressure: the effective planetary radii, the atmospheric
  temperature profiles, and the abundances of several species.  We
  discuss our models in Section~\ref{sec:models}, and show
  transmission spectra in Figure~\ref{fig:modelspec}
  and~\ref{fig:modelspec_zoom}.}

For each model we compute the monochromatic planetary radii,
$R_P(\lambda)$, with PHOENIX under the assumption of spherical
symmetry \citep{barman:2007} and with a wavelength sampling of
0.05~\AA\ from $1-3$~\micron.  Each radius spectrum is then convolved
with a Gaussian kernel to the instrumental resolution and then
interpolated onto the observed wavelengths.  We also compute the
quantity $\sigma_{R_P}$, defined as the standard deviation of the
planet radius at our model resolution over the wavelength range used
in our analysis.  A completely flat spectrum would have
$\sigma_{R_P}=0$ and be undetectable by our methods, but models with
sufficiently prominent spectral features -- which we can detect --
have larger values.  To facilitate comparison of $\sigma_{R_P}$ with
our sensitivity limits we list these values in Table~\ref{tab:results}
with the final results of our analysis.

Note that all of our models are H$_2$/He-dominated.  There are many
possible alternative compositions
\citep{miller-ricci:2010,rogers:2010}, but the atmospheric scale
heights of those dominated by heavier molecules result in suppressed
transmission spectra with features undetectable at our precision.  We
list some of the models we test, and the $R_P(\lambda)$ they predict
when observed with CFHT/WIRCam with the J, CH4-on, and Ks filters, and
with Spitzer/IRAC CH1 (3.6~\micron) and CH2 (4.5~\micron), in
Table~\ref{tab:models}.  We plot $R_P(\lambda)$ from a selection of
our models in Figure~\ref{fig:modelspec} across the near-infrared, and
in Figure~\ref{fig:modelspec_zoom} zoomed in on the spectral regions
probed by our NIRSPEC observations.

\figtwocol{modelspec}{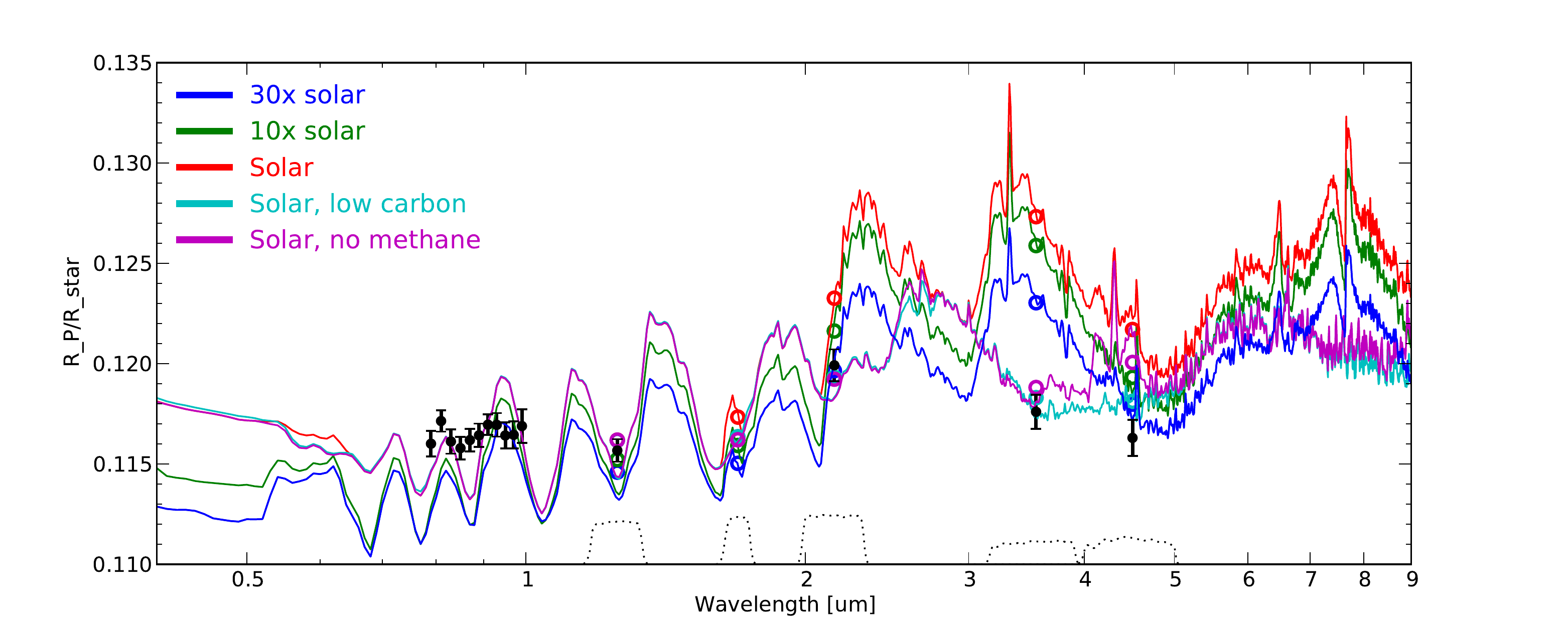}{width=7in}{}{A selection of
  our model transmission spectra of \gjb\ (solid lines; smoothed for
  display purposes).  The dashed lines at bottom represents the
  effective throughput of the WIRCam J, CH4-on, and Ks and IRAC 1 \& 2
  filters. The open circles show the averages of our models over the
  several filters, and the solid points are the observations
  \citep{bean:2010,desert:2011,croll:2011b}.  The spectroscopic
  analysis presented in this paper rules out all but the low carbon
  and no methane models (cf. Table~\ref{tab:results}). We describe
  these models in Section~\ref{sec:models} and
  Table~\ref{tab:models}. }

\figtwocol{modelspec_zoom}{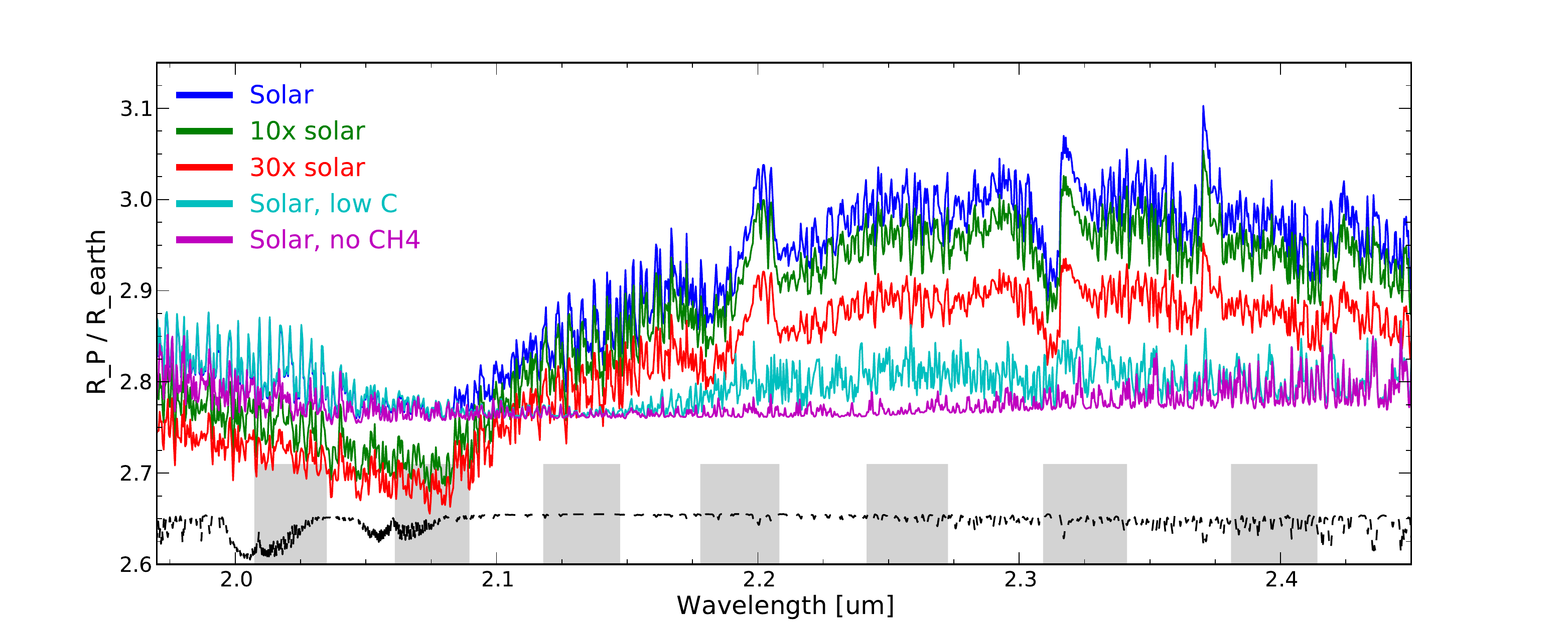}{width=7in}{}{A
  selection of the same model transmission spectra shown in
  Figure~\ref{fig:modelspec}, showing only the wavelength range
  covered by our observations and smoothed for display purposes. The
  shaded regions represents those wavelength we observed; the regions
  are discontinuous due to NIRSPEC's echelle format.  The dashed line
  represents the telluric transmission \citep{hinkle:2003}.}

\begin{deluxetable}{l  c c c c c c}
  \tablecolumns{5} \tablecaption{\label{tab:models} \gjb\ Predicted
    Photometric Transit Radii (cf. Figure~\ref{fig:modelspec}) }

\tablewidth{0pt}
\tablehead{
\colhead{Model}  &  \colhead{$R_{min}$\tablenotemark{a}/ \rearth}  &  \colhead{J\tablenotemark{b}}  &  \colhead{CH4-on\tablenotemark{b}}  &  \colhead{Ks\tablenotemark{b}}  &  \colhead{IRAC1 (3.6~\micron)}  &  \colhead{IRAC2 (4.5~\micron)}}

\startdata 
Solar  &  -   &    2.665  &  2.692  &  2.828  &  2.921  &  2.792  \\ 
Solar  & 2.85 &    2.850  &  2.850  &  2.877  &  2.921  &  2.851  \\ 
Solar  & 2.90 &    2.900  &  2.900  &  2.909  &  2.931  &  2.900  \\ 

10x solar  &  -   &    2.643  &  2.659  &  2.790  &  2.888  &  2.737  \\ 
10x solar  & 2.85 &    2.850  &  2.850  &  2.865  &  2.892  &  2.850  \\ 
10x solar  & 2.90 &    2.900  &  2.900  &  2.901  &  2.911  &  2.900  \\ 

30x solar  &  -   &    2.629  &  2.639  &  2.745  &  2.823  &  2.707  \\ 

Solar, low carbon  &  -   &    2.665  &  2.669  &  2.737  &  2.714  &  2.710  \\ 

Solar, no methane  &  -   &    2.665  &  2.666  &  2.735  &  2.726  &  2.754  \\ 

\enddata
\tablenotetext{a}{Planetocentric radius of artificially imposed near-infrared opaque
  cloud deck; cf. Section~\ref{sec:models}}
\tablenotetext{b}{Near-infrared radii computed using the CFHT/WIRCam
  filter profiles at
  \url{http://www.cfht.hawaii.edu/Instruments/Filters/wircam.html}}.
\end{deluxetable}

In the infrared, as in the optical, clouds can alter the transit
spectrum by blocking flux below a certain radius, establishing
different minimum radii compared to cloud-free models.  The impact of
cloud opacity was tested for the cloud-free models described above by
truncating the model radius spectrum at several different values, in
effect inserting an opaque cloud deck at a desired level in the
atmosphere.  This means that in our cloudy cases we set all values of
$R_P(\lambda)$ less than a given radius equal to that radius.

\section{NIRSPEC Observations and Analysis}
\label{sec:obsanal}
\subsection{Summary of Observations}

The temperature of \gjb\ is such that we expect
 many prominent spectral features in the near infrared, and we therefore 
observed \gjb\ using the NIRSPEC cryogenic near-infrared echelle
spectrograph \citep{mclean:1998}, located at the Nasymth focus of the
Keck II telescope.  Our observations covered one transit each in both
the H and K near-infrared bands; we list the details of these
observations and our instrumental setup in
Table~\ref{tab:observations}.

\begin{deluxetable}{l c c }
\tablecolumns{3}
\tablecaption{\label{tab:observations} Observations }
\tablehead{                   & 2010 Aug 16  & 2010 Sep 04 }
\tablewidth{0pt}
\startdata
Cross-disperser position  & 35.60 & 36.89  \\
Echelle position          & 63.82 & 62.94 \\
Slit                     & 0.72'' x 24'' & 0.72'' x 12''  \\
Filter                   & NIRSPEC-7 & NIRSPEC-5 \\
Wavelength coverage (\micron)   &   2.007 - 2.414 & 1.514 - 1.783 \\
Spectral grasp &  52\% & 70\% \\
Integration Time (sec)   & 60 & 60 \\
Number of co-adds        & 1 & 1 \\
Number of exposures      & 182 & 85 \\
Airmass range            & 1.04 - 1.65 & 1.15 - 1.65 \\
\gjb\ phase coverage           & -0.060 - +0.038 & -0.021 - +0.025\\
\enddata
\end{deluxetable}    

On both nights we observed the \gj\ system continuously for as long as
conditions permitted.  We designed our observations to minimize
possible sources of systematic error, and thus did not nod the
telescope.  Our prior experience with NIRSPEC and with SpeX (a
moderate-resolution spectrograph at the NASA Infrared Telescope
Facility) has taught us that nodding induces an undesirable
``sawtooth'' pattern in the extracted spectrophotometry (presumably
due to residual flat-field variations), while it does not provide a
substantial improvement in removing background emission. Instead, we
kept the target star in a fixed position about 6'' from the end of the
spectrograph slit; this allowed us to squeeze an extra echelle order
onto the detector. For the same reason, we deactivated the
instrument's field rotator; differential atmospheric refraction did
not appear to affect the relative tilt of our spectra over the course
of the night.  We used the widest possible slit (0.7'') to maximize
throughput and minimize the effects of sub-arcsecond guiding errors.
Typical frames had maximum count rates of roughly 5,000~e$^-$ per
pixel, safely within the ALADDIN-3 detector's linear response range.
We chose not to switch repeatedly between target and calibrator stars
because (a) of the limited amount of observation time (\gjb 's transit
only lasts about an hour), and (b) in our experience each
re-acquisition of the target can induce substantial flux
discontinuities in spectrophotometric time series.  Although we did
acquire a few observations of a telluric calibrator star (the A0V star
HD~161289) at high airmass we do not use these in our analysis: due to
the nature of our analysis dividing our spectra of GJ~1214 by a
single, constant telluric calibrator spectrum (as is typically done
for such observations) will add noise into our data but will not
otherwise affect our final results.

\subsection{Initial Data Reduction}
\label{sec:reduction} We reduce the raw echelleograms using a
combination of standard IRAF routines (as implemented in
PyRAF\footnote{PyRAF is a command language for running IRAF tasks that
  is based on the Python scripting language, and is available at
  \url{http://www.stsci.edu/resources/software_hardware/pyraf/}}) and
our own set of Python analysis tools\footnote{Available at
  \url{http://www.astro.ucla.edu/~ianc/}}; we also draw on the
experience of similar past observations \citep[e.g.,][]{deming:2005}.
We dark-subtract the frames and interpolate over cosmic rays and bad
pixels.  We performed flat-field correction for our data using the
NIRSPEC internal calibration lamp, then using the IRAF task
\texttt{apnormalize} to remove the lamp's intrinsic continuum and
correct small-scale flat-field variations.

The two left-hand (shorter-wavelength) quadrants of the NIRSPEC
detector sometimes exhibit nonuniform readout characteristics in a
pattern that repeats every eight detector rows. This manifests itself
both as increased read noise and an apparent bias offset in affected
rows. We remove the offset from each set of rows in each quadrant
using linear least squares after excluding pixel values more than
seven standard deviations from the median. Nonetheless the
short-wavelength half of each extracted echelle order is detectably
noisier.

We extract the spectra using the IRAF \texttt{apall} task with
variance weighting and a third-order sky background fit.  Our
extraction ignores the slight tilt of the spectrograph slit in the raw
spectral orders, but this does not seriously compromise our resolution
(R$\approx$17,000) when using the 0.7'' slit. The extracted spectra
have fluxes of roughly 16,000 e$^-$~pix$^{-1}$.  After removing
observations rendered unusable for telescope or instrumental reasons
(e.g., loss of guiding or server crashes), we are left with 182 K-band
and 85 H-band spectra.

Our initial analysis reveals that the the H-band data are unable to
constrain the presence of even the most easily detectable atmospheric
models. The cause seems to be the relatively short duration of the H
band data (cf. Table~\ref{tab:observations}), and especially the
paucity of out-of-transit observations.  We repeated the analysis we
describe below using only 80 of the 182 K-band frames (similar in
length to the H band data set); under these conditions we achieve
precision comparable to that from the H band dataset. Therefore it
seems likely that the limiting factor in the H band data set is an
insufficient number of observations and/or insufficient pre- and
post-transit coverage.  We thus discard the H band data and focus
solely on the K band data.

We calculate a wavelength solution for our spectra using the IRAF task
\texttt{ecidentify} to identify isolated telluric absorption lines in
each extracted spectrum.  As our reference we use the high-resolution
telluric absorption spectrum measured by \cite{hinkle:2003}.  We fit a
third-order dispersion solution to all echelle orders simultaneously;
our final wavelength solution has RMS residuals of 0.3~\AA\ (0.7
pixels).  The largest residual we see is 0.6~\AA\ (2.2 pixels) in the
2.02\micron\ echelle order, but as this order is heavily corrupted by
absorption from telluric \coo\ we do not use data from this order in
our final analysis: the largest residual we see that could directly
affect our data is 0.45\AA\ (1.5~pixels) in the 2.13\micron\ echelle
order.

To convert observed wavelengths to wavelengths in the \gj\ system's
rest frame we use a radial velocity of 48~\kms, which we estimate by
cross-correlating a high-resolution model stellar spectrum with our
spectra.  Before cross-correlation we used the IDL package XTELLCORR
\citep{vacca:2003} to correct our \gj\ spectra for telluric absorption
using observations we took of the A0V star HD~161289.  WIth the IDL
routine \texttt{BARYVEL} we estimate a solar system barycentric
correction of $-23.8$~\kms\ for the date of our K band observations;
the resulting systemic velocity of 24~\kms\ is in decent agreement
with the value of 21.0~\kms\ from \cite{berta:2011}.  Finally, we
convert the UTC timestamps in our NIRSPEC files to heliocentric Julian
dates (HJD) using the function \texttt{helio\_jd} in the Python
\texttt{astrolib} module.

\subsection{Measuring systematic effects}
Our spectra, shown in Figure~\ref{fig:rawdata}, exhibit substantial
residual temporal variations due to a combination of instrumental,
telluric, and astrophysical sources (i.e., the planetary transit),
with the last the weakest of these effects.  We wish to quantify and
ultimately remove the instrumental and telluric effects to the extent
that we can convincingly detect any wavelength-dependent variations in
the transit depth.

\figtwocol{rawdata}{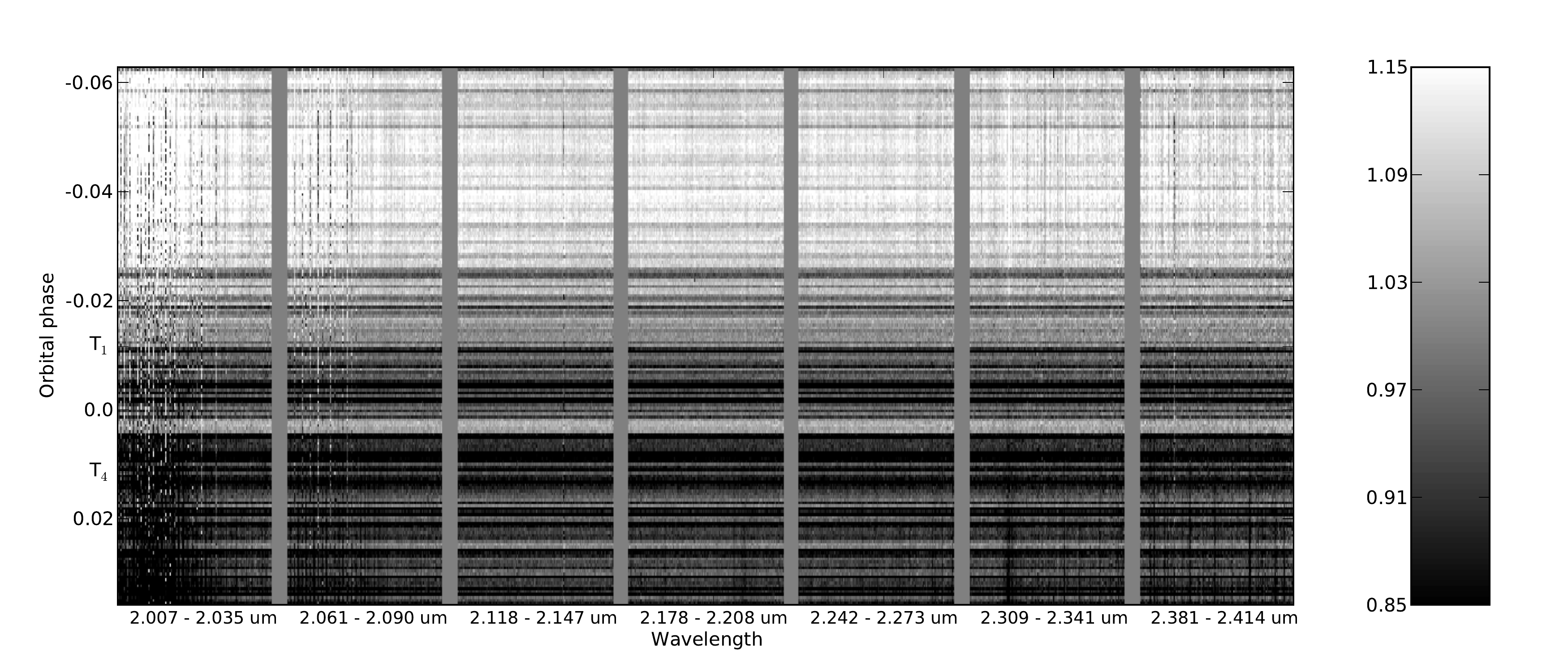}{width=7in}{}{Spectrophotometric data
  $f_{i\lambda}$, normalized by the median value in each wavelength
  channel.  The variations (due to a combination of airmass effects
  and instrumental slit loss) are of order 10\% (which prevents us
  from directly seeing \gjb 's transit) and are largely common mode.
  For display purposes the data from each echelle order are separated
  by a narrow gray region. $T_1$ and $T_4$ denote the beginning of
  transit ingress and the end of egress \citep{carter:2011}. }

The widest slit available on NIRSPEC is only 0.7'' across, so even in
excellent seeing significant light is lost and does not enter the
slit.  In 0.6'' seeing (about the best we observed), guiding errors of
0.1'' can cause the amount of light entering the spectrograph to vary
by as much as 10\% from one frame to the next.  We see variations at
this level in our spectrophotometric data, which we attribute to
pointing jitter.  These variations wholly overwhelm the 1.4\%\ flux
decrement from \gjb 's transit.

We quantify the amount of light coupled into the spectrograph slit by
measuring the flux in spectral regions free of telluric absorption
lines; we avoid these lines by referring to a high-resolution telluric
absorption spectrum \citep{hinkle:2003}.  We list the spectral regions
we use in Table~\ref{tab:telluric}.  The flux in these channels should
only depend on (a) frame-to-frame changes in starlight entering the
spectrograph slit, and (b) atmospheric continuum extinction.  In each
spectrum we add up the flux in these telluric-free regions, creating a
time series representative of the slit losses suffered by the
instrument.  We denote this quantity as $\ell$ and refer to it as the
slit loss, though it is actually a combination of instrumental slit
losses and telluric continuum absorption due to changing airmass.  The
slit loss is plotted in Figure~\ref{fig:phot} with other (cleaned)
spectrophotometric time series, and in Figure~\ref{fig:state_vectors}
along with other observable systematics.

\figtwocol{phot}{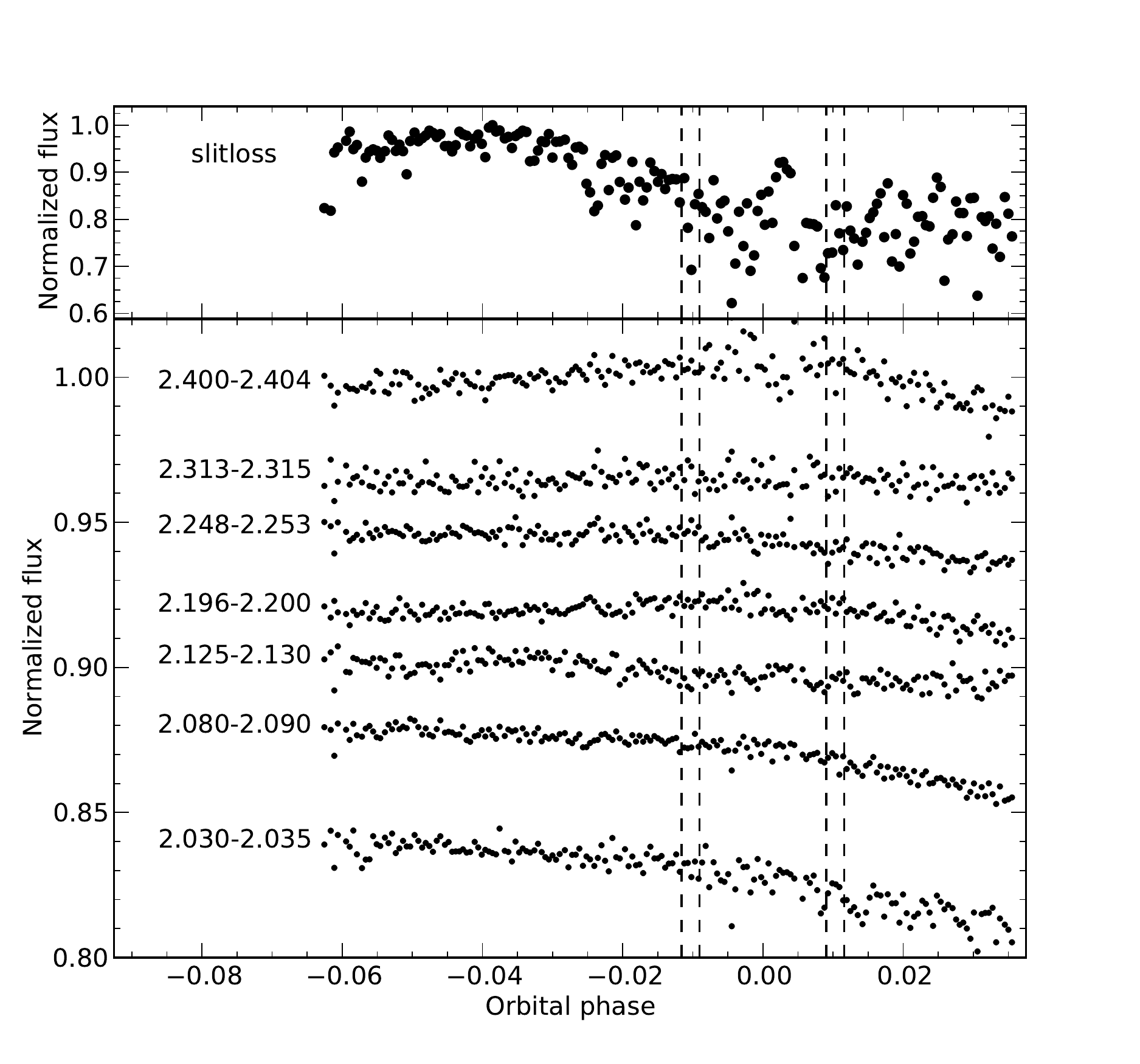}{width=7in}{}{Representative
  spectrophotometric time series.  The top panel shows the relative
  flux coupled into the spectrograph slit, as measured in regions free
  of telluric absorption lines; telluric continuum absorption and
  guiding errors combine to produce 10\% variations, masking the 1.4\%
  transit signature.  The bottom panel shows one time series from each
  echelle order, after removal of the common mode slit loss term and
  binned over the wavelength range listed (in \micron).  After
  correcting for airmass effects these data have variations of order
  0.5\%, but the transit is still not visible: this is because
  dividing out the common-mode slit loss term removes a mean transit
  profile from all the data. The dashed lines indicate the times of
  first through fourth contact of the transit \citep{carter:2011}.}

\figtwocol{state_vectors}{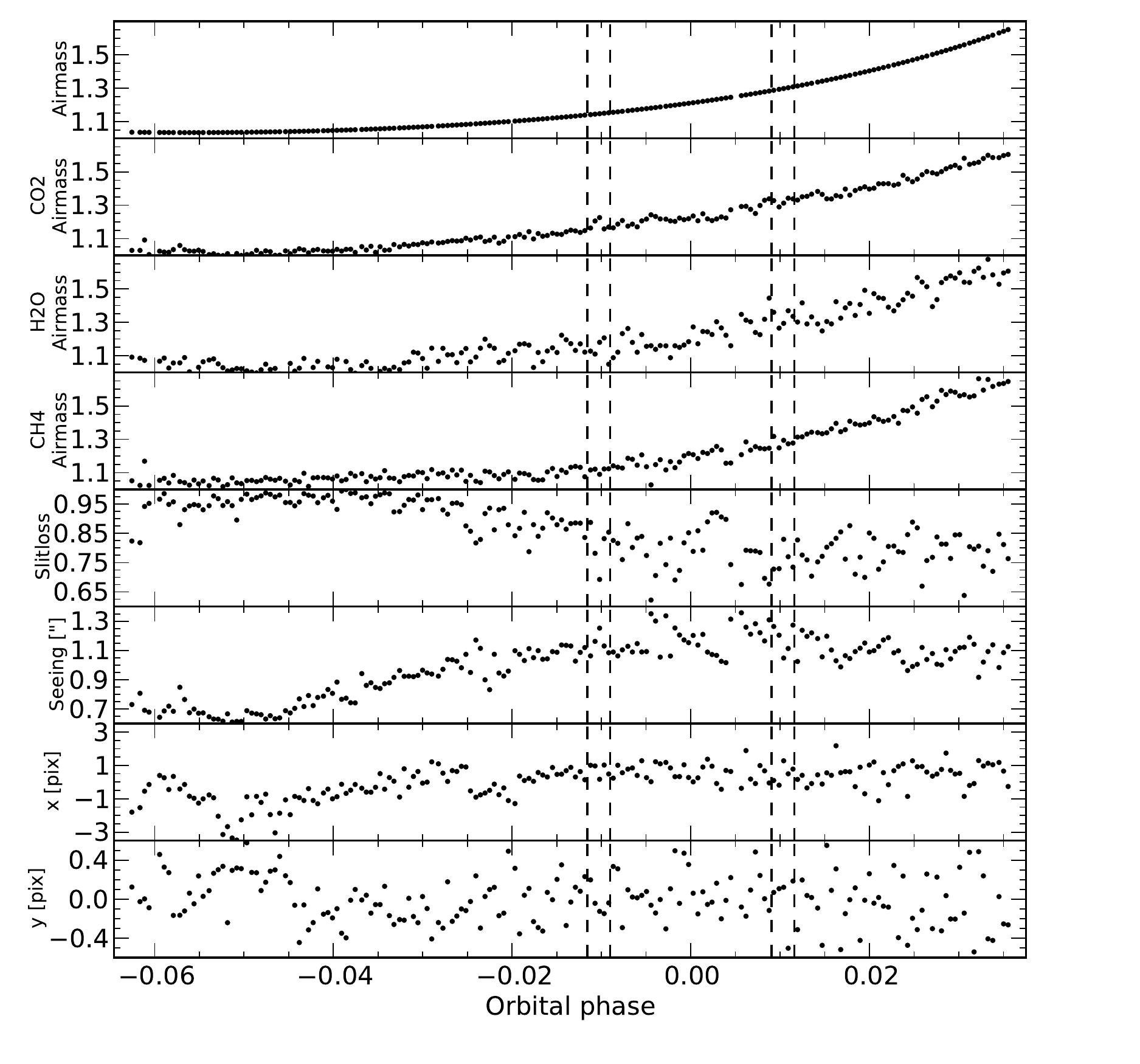}{width=7in}{}{The various
  observable quantities (described in Section~\ref{sec:reduction})
  measured during the course of our observations.  As described in the
  text, we ultimately fit our data using the slitloss term, an
  empirical airmass quantity derived from the telluric \coo\ line
  flux, and several low-order polynomials. The dashed lines indicate
  the times of first through fourth contact of the transit
  \citep{carter:2011}.}

\begin{deluxetable}{c }
\tabletypesize{\scriptsize}
\tablecolumns{1}
\tablewidth{0pt}
\tablecaption{\label{tab:telluric} Telluric-free spectral regions from \cite{hinkle:2003}  }

\tablehead{ \colhead{Wavelength range (\micron)}  }

\startdata

2.1264290 -  2.1265542  \\ 
2.1297217 - 2.1299722  \\ 
2.1304068 - 2.1305100  \\ 
2.1313866 - 2.1317770  \\ 
2.1320127 - 2.1322632  \\ 
2.1326462 - 2.1327346  \\ 
2.1332503 - 2.1333681  \\ 
2.1334270 - 2.1336333  \\ 
2.1344289 - 2.1349077  \\ 
2.1350624 - 2.1355338  \\ 
2.1364841 - 2.1367567  \\ 
2.1370218 - 2.1371176  \\ 
2.1378911 - 2.1382520  \\ 
2.1386277 - 2.1388855  \\ 
2.1390697 - 2.1391655  \\ 
2.1397769 - 2.1398579  \\ 
2.1400789 - 2.1409260  \\ 
2.1410071 - 2.1413459  \\ 
2.1414859 - 2.1418321  \\ 
2.1438800 - 2.1439978  \\ 
2.1798910 - 2.1801862  \\ 
2.1804436 - 2.1808296  \\ 
2.1809886 - 2.1824343  \\ 
2.1856588 - 2.1869077  \\ 
2.1873467 - 2.1874754  \\ 
2.1884518 - 2.1885654  \\ 
2.1886411 - 2.1887849  \\ 
2.1889590 - 2.1891407  \\ 
2.1898219 - 2.1902609  \\ 
2.1907680 - 2.1924106  \\ 
2.1924938 - 2.1927512  \\ 
2.1928723 - 2.1940153  \\ 
2.1940758 - 2.1943105  \\ 
2.1953777 - 2.1955367  \\ 
2.1959757 - 2.1961422  \\ 
2.2077762 - 2.2078973  \\ 
2.2447114 - 2.2450073  \\ 
2.3109368 - 2.3111936  \\ 
2.3125575 - 2.3127902  \\ 
2.3128624 - 2.3129587  \\ 
2.3132796 - 2.3134401  \\ 
2.3138813 - 2.3140418  \\ 
2.3140980 - 2.3145473  \\ 
2.3147318 - 2.3148201  \\ 
           
\enddata   
\end{deluxetable}

High-precision near-infrared photometry demonstrates that apparent
photometric variations can be induced by motion of a star across the
detector, changes in seeing, and other instrumental sources
\citep[e.g.,][]{rogers:2009}.  We expect the same to be true for
high-precision spectrophotometry, and so we measure the motion of the
spectral profiles in the raw echelleograms parallel to ($x$) and
perpendicular to ($y$) the long axis of the spectrograph slit.  We
also measure the total amount of light coupled into the spectrograph
slit, and the FWHM of the seeing profile.  All these quantities are
plotted in Figure~\ref{fig:state_vectors}.

We measure the $x$ motion of the spectra on the detector by
spline-interpolating the spectrum in each echelle order and
cross-correlating it at sub-pixel increments with a high signal to
noise (S/N) template spectrum \citep{deming:2005}.  We construct our
template by taking the temporal average, after removing outliers, of
all our spectra.  A parabolic fit to the peak of each spectrum's cross
correlation provides the optimal offset value, and we then
spline-interpolate all the spectra to the template's reference frame.

We measure the $y$ motion of the star along the slit axis by tracing
the locations of the spectral profiles using the IRAF \texttt{apall}
task. The locations of the spectra on the detector are not
significantly affected by differential atmospheric refraction, so we
use the same set of relative positions for data at all wavelengths.
Both $x$ and $y$ motions are plotted in
Figure~\ref{fig:state_vectors}; though we see evidence that these
motions affect our spectrophotometry (as we discuss in
Section~\ref{sec:sysrem} below), this is a low-amplitude effect.

Though we do not have a measure of the true astronomical seeing during
the night, the width of the spectral profiles in the raw echelleograms
provides a proxy for seeing.  We fit a cubic spline to the mean
spectral profile for each echelle order, and measure the average full
width at half maximum in each frame.  This quantity depends on a
combination of astronomical seeing, instrumental focus, and pointing
jitter during an exposure, but we hereafter refer to it simply as
seeing.

Previous studies \citep[e.g.,][]{deming:2005} report that an empirical
measure of atmospheric absorption is preferable to the airmass value
calculated from the telescope's zenith angle when accounting for
telluric extinction.  To test this, we measure the flux in sets of
telluric absorption lines caused by CO$_2$, \methane , and H$_2$O.  We
select these lines, and verify that each is due to only a single
absorber, by generating synthetic single-species telluric absorption
spectra using the ATRAN code \citep{lord:1992}.  We fit the
slitloss-corrected flux, $f'_i$, in these single-species telluric time
series to the function $\ln f'_i = p + q a_i$, where $a_i$ is the
airmass at timestep $i$. We then define each species' empirical
airmass term as $a'_i = \left( \ln f'_i - p \right) / q$.  These
different airmass terms are all plotted in
Figure~\ref{fig:state_vectors}.

\subsection{Identifying and Removing Systematic Effects}
\label{sec:sysrem} After extracting the spectra and quantifying observable
systematic effects, our next step is to remove the large-scale flux
variations caused by slit losses and (continuum) telluric absorption
(note that this is the point in the analysis at which we inject
synthetic model transit signals, as discussed later in
Section~\ref{sec:injection}). We divide the flux in every wavelength
channel by the slit loss time series, viz.: $f'_{i\lambda} =
f_{i\lambda}/\ell_\lambda$.  This improves the S/N in regions clear of
telluric absorption lines from roughly 10~pixel$^{-1}$ to
roughly 50~pixel$^{-1}$ (cf. Figures~\ref{fig:rawdata}
and~\ref{fig:corrdata}) -- this is still $>3$~times worse than
expected from photon noise. This step removes a mean transit profile
from all wavelength channels, but the overall shape of the
transmission spectrum should remain the same.  We show a set of
binned, slitloss-corrected spectrophotometric time series in
Figure~\ref{fig:phot}, and show the full set of $f'_{i\lambda}$ in
Figure~\ref{fig:corrdata}.

\figtwocol{corrdata}{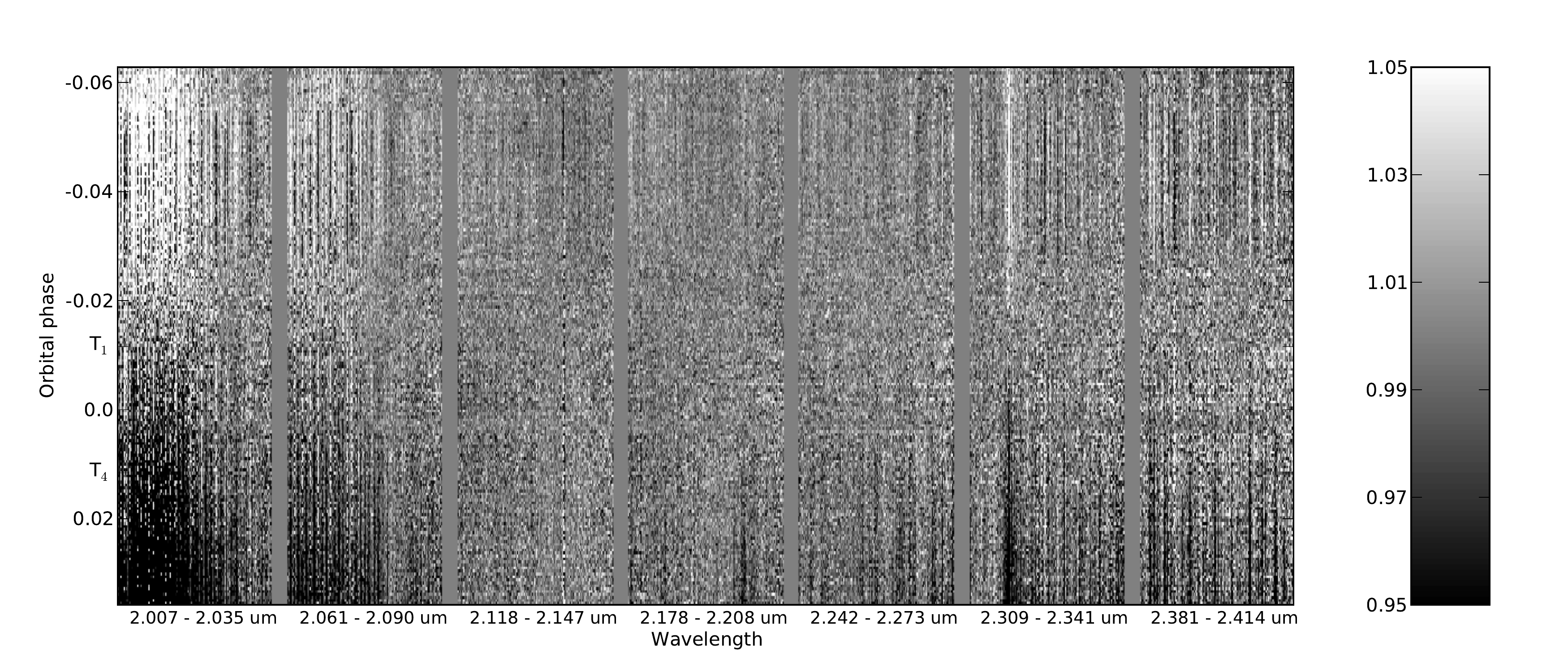}{width=7in}{}{Spectrophotometric
  data $f'_{i\lambda}$ after dividing all wavelength channels by the
  common mode slit loss term (shown in Figure~\ref{fig:phot}), and
  normalized by the median value in each wavelength channel.  The data
  shown have variations of order 0.5\% after correcting for residual
  airmass trends, but still no transit is visible because dividing by
  the slit loss term has removed the mean transit depth from all
  wavelength channels.  $T_1$ and $T_4$ denote the beginning of
  transit ingress and the end of egress \citep{carter:2011}. For
  display purposes the the data from each echelle order are separated
  by a narrow gray region. }

The most obvious remaining variability is correlated with airmass.  We
identify other possible sources of variability using Principal
Component Analysis \citep{joliffe:1986}.  We compute the strongest
several principal components (PCs) in our slit loss-corrected data
set, and look for correlations between these PCs and our systematic
observables: i.e., $x$ and $y$ motions, seeing, slit losses, sky line
emission, and our several airmass proxies.  We observe correlations
between strong principal components and airmass, $x$, $y$, and slit
losses, but not with other instrumental parameters.  We attribute the
$x$ and $y$ dependence to residual flat-field variations as each
resolution element moves on the detector.  The correlation of slit
losses (which have already been removed to first order) with a
principal component which is most prominent in regions of strong
telluric absorption suggests flux variations in the core of telluric
lines not well explained by a simple exponential extinction relation.
As described in Section~\ref{sec:fitting}, we ultimately detrend our
data using only the slit loss term because our precision is not high
enough to justify using additional correction terms..

\section{Needle in a Haystack: Searching for Transit Signatures}
\label{sec:detection} We are unable to measure \gjb 's transit light curve
directly, because this signal is buried in the slit loss term we
remove from all spectral time series.  This removal has the effect of
subtracting a constant offset value $C$ from the transmission spectrum.
The overall shape of the spectrum is therefore unchanged, and we can
still hope to measure the modified transmission spectrum -- i.e.,
$d_\lambda = (R_P/R_*)^2 - C$.

If \gjb\ has a cloudless, H$_2$/He-dominated atmosphere we expect
substantial variations in $d_\lambda$ across the wavelength ranges we
observe \citep[cf. Figure~\ref{fig:modelspec};
also][]{miller-ricci:2010}; but if the planet has a denser atmosphere
(and thus a smaller atmospheric scale height) we expect to see an
essentially flat transit spectrum -- in this case we might see no
transit signature in any wavelength channel.  Clouds would further
suppress any spectral features occuring below a given altitude.

Photon noise alone precludes detection of the residual differential
transit signal in any single spectral line.  Even when examining time
series that average over broad spectral regions (e.g., molecular
bandheads) we are unable to detect any transit-like features.  This
agrees with our conclusion that, at our sensitivity, the spectrum is
consistent with being featureless. To achieve our final sensitivity we
must use our entire wavelength coverage.  Our data do not permit us to
search for any individual spectral features
\citep[cf.][]{redfield:2008}, but they do allow us to test a given
model against the entirety of our data.

Note that we cannot use exactly the same cross-correlation technique
successfully used by \cite{snellen:2010} in their detection of CO on
HD~189733b.  This is because (a) \gjb\ exhibits a much smaller change
in velocity (12 km~s$^{-1}$) during a transit than does HD~189733b,
(b) our spectral resolution is much lower (17,000 with NIRSPEC
vs.~100,000 with CRIRES), and (c) the signal to noise of our (fewer)
spectra are substantially lower.  For similar reasons, and because the
most optimistic models of planetary emission predict planet-to-star
contrast ratios of $\lesssim10^{-4}$ for \gjb, we cannot use the
spectral deconvolution techniques used by \cite{barnes:2010} to search
for planetary emission.

Instead, we search for residual transit signatures in our data by
fitting a model (including telluric, systematic, and transit effects)
to the time series in each wavelength channel, as described in
Section~\ref{sec:fitting}.  We then compare the extracted transmission
spectrum to model spectra using both cross-correlation and linear
least squares techniques, as described in Section~\ref{sec:comparing}.

\subsection{Fitting to the data}
\label{sec:fitting} We fit each spectral time series (i.e., the
slitloss-corrected flux in each wavelength bin) with the following
relation:
\begin{equation}
f'_{i\lambda} = f_{0\lambda}  e^{b_\lambda {a_i}}  \left( 1 + d_\lambda r_i \right) 
  \left( 1 + \sum_{j=1}^J c_{j\lambda}v_{ji} \right)  
\label{eq:fluxeqn}
\end{equation}  
This equation represents a transit light curve affected by systematic
and telluric effects.  In all cases the subscript $i$ refers to the
frame number and $\lambda$ to the particular wavelength bin.  The
variables are: $f'_{i\lambda}$, the slitloss-corrected measured flux;
$f_{0\lambda}$, the out-of-transit flux that would be measured above
the Earth's atmosphere; $r_i$, the flux in a transit light curve
scaled to equal zero out of transit and approximately $-1$ inside
transit \citep[i.e., $(F - 1) / (R_P/R_*)^2$, using the nomenclature
of][]{mandel:2002}; $d_\lambda$, the depth of transit; $v_{ij}$, the
$J$ state vectors expected to have a linearly perturbative effect on
the instrumental sensitivity; $c_{j\lambda}$, the coefficients for
each state vector; and $a_i$, the airmass, which has an exponential
extinction effect on the measured flux modulated by an extinction
coefficient $b_\lambda$.  Since residual, low-amplitude drifts are
common in this kind of observation \citep[cf.][]{swain:2010,bean:2010}
we account for and remove the effect of these drifts by including a
series of polynomials in the set of vectors $v_{ji}$. For our
polynomials we use the Chebychev polynomials of the first kind, which
we compute as functions of orbital phase normalized to the domain
$[-1, +1]$ to preserve the orthogonality of the set; as we describe
below, we ultimately use the first three (linear through cubic)
Chebychev polynomials in our analysis.  After fitting we obtain a set
of coefficients $(f_{0\lambda}, b_\lambda, c_{j\lambda}, d_\lambda )$
from our full set of observations; the $d_\lambda$ represent our
measured transmission spectrum.

We use the parameters of \cite{carter:2011} for our model transit
light curve and use a nonlinear limb-darkened light curve using the
small-planet approximation from \cite{mandel:2002}.  We take our
near-infrared limb-darkening coefficients from \cite{claret:2000} for
a star with $T=3000$~K and surface gravity $10^5$~cm~s$^{-2}$. In any
case we are not especially sensitive to different limb-darkening
models -- even a uniform-disk transit curve leaves our results
unchanged within their uncertainties.

We initially fit our data using many combinations of state vectors and
low-order polynomials, and both calculated and empirical airmass
quantities.  We use the Bayesian Information
Criterion\footnote{Bayesian Information Criterion (BIC) = $\chi^2 + k
  \ln N$, where $k$ is the number of free parameters and $N$ the
  number of datapoints.}  (BIC) to choose which of these many
parameter sets best fit our data.  Calculating the BIC involves
computing $\chi^2$ from each set of parameters, which in turn requires
us to assign uncertainties to each datapoint.  We estimate these
uncertainties as follows.  We first compute unweighted fits of
Eq.\ref{eq:fluxeqn} (using no $v_{ji}$, and airmass as reported from
the telescope control system) to the data using a multivariate
minimization provided in the SciPy\footnote{Available at
  \url{http://www.scipy.org/}.}  software distribution (the function
\texttt{optimize.leastsq}), and compute the residuals for each time
series.  We then assign uncertainties to the individual datapoints by
adding in quadrature the expected photon noise and the standard
deviation of the residuals in each time series; this weights more
heavily frames with the highest throughput, which are presumably those
least affected by systematically lower telluric transmission and/or
slit loss from guiding errors.  Our next step is to globally scale
down all uncertainties in each wavelength channel to give a $\chi^2$
equal to the number of datapoints.  We then use these scaled values as
our per-point measurement uncertainties.  Although this method of
estimating uncertainties is not statistically valid inasmuch as it
overestimates our goodness-of-fit (because we have artificially
reduced $\chi^2$; remember also that that even near-infrared
photometry typically only comes within a factor of 3-5 of the photon
noise limit) and underestimates parameter uncertainties
\citep{andrae:2010}, it provides a quantitative measure with which to
compare the relative merit of various models.

The instrumental model which gives the lowest global BIC for our data
include polynomials up to cubic order, the slit loss term described
previously, and an empirical airmass calculated from the sum of flux
in unsaturated telluric \coo\ lines from $2.0-2.1$~\micron.  We confirmed
that this result is not dominated by the first two echelle orders
(populated with the numerous telluric \coo\ lines from which we
construct our empirical \coo\ airmass) by excluding these orders and
repeating the BIC analysis; the same set of parameters still deliver
the lowest BIC.  

The relative transit depth spectrum $d_\lambda$ that results from the
fitting process represents a modified transmission spectrum of \gjb ;
specifically, $d_\lambda = (R_P(\lambda)/R_*)^2 - C$.  The constant
$C$ is the mean transit depth removed by our correction for slit loss
variability.  For display purposes, we show our extracted spectrum --
averaged in wavelength -- in Figure~\ref{fig:extracted_spec}; note
that we do not fit to the binned data shown there.  Regions of strong
telluric absorption (especially for $\lambda < 2.1$~\micron) are
notably biased, and we see a distinct tilt in all echelle orders.  We
suspect (but cannot independently confirm) that this tilt is some sort
of detector artifact; we also see it in the spectra we extract after
injecting model spectra into our data
(cf. Section~\ref{sec:injection}).  We do not know the origin of this
intra-order slope; though because we make an inherently relative
transit fit in each wavelength channel we do not believe it could be
due to spatial detector sensitivity variations.  Removing this tilt
from each order does not markedly increase our sensitivity, and to be
conservative we do not remove it from our data. Our sensitivity tests
demonstrate that we are still sensitive to plausible injected models
in the presence of this slope.

\figtwocol{extracted_spec}{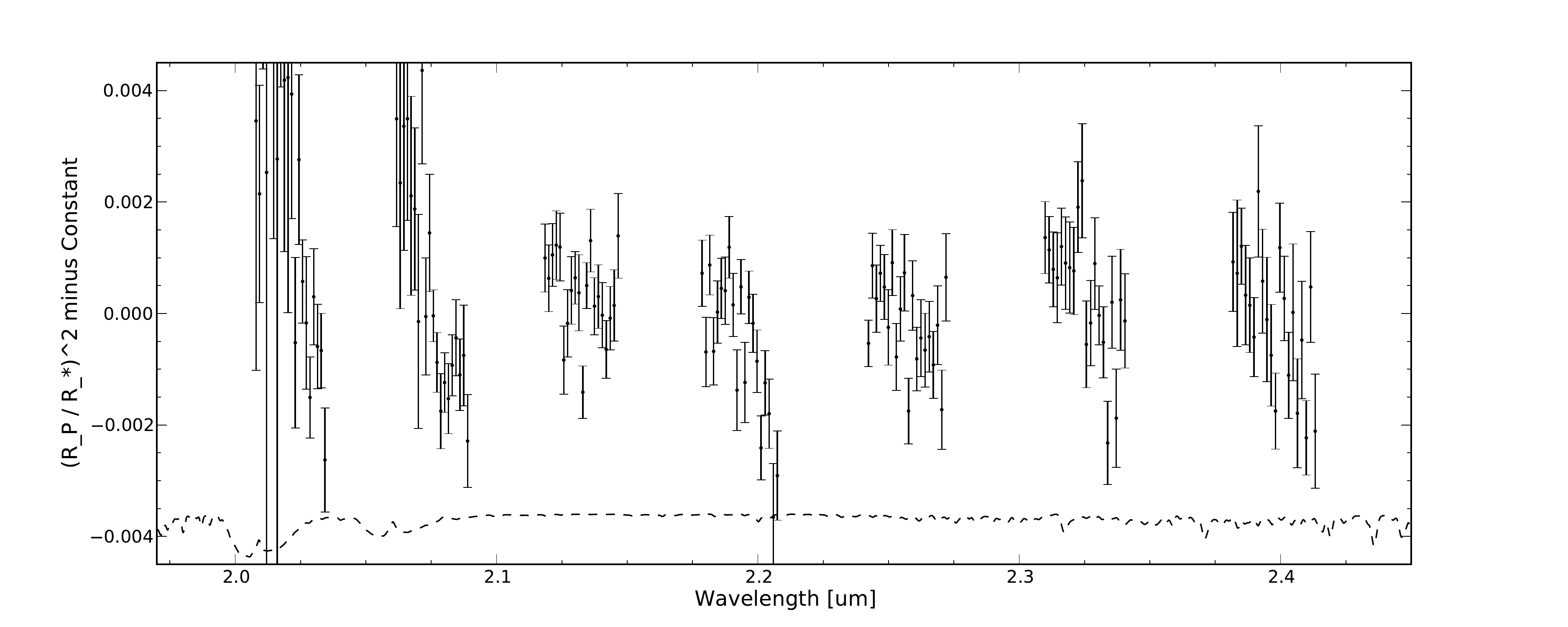}{width=7in}{}{Extracted
  transmission spectrum of \gjb , binned to 1.5~nm for display
  purposes; error bars here represent solely statistical (standard
  deviation on the mean) errors.  Regions of strong telluric
  absorption (e.g., shortward of 2.1~\micron) are notably biased.  We
  do not know the source of the tilt in each echelle order. The dashed
  line at bottom represents the telluric transmission from
  \cite{hinkle:2003}.}

We assess $\sigma_{d_\lambda}$, the uncertainties on $d_\lambda$, in
two ways: using the residual permutation (``prayer-bead'') method
\citep{gillon:2007}, and by quantifying how our extracted parameters
vary as we inject model transits at different times.  The residual
permutation technique fits fake datasets that are generated by adding
temporally shifted copies of the residuals to the best-fit model.
Thus it is similar to the bootstrap method, but has the advantage of
preserving the properties of correlated noise (present in our data).
We estimate the 68.3\% confidence interval on the $d_\lambda$ by
marginalizing over all other parameters and determining the parameter
values that enclose the central 68.3\% of the resulting
one-dimensional parameter distributions.

Our second technique to determine parameter uncertainties injects
model transit spectra into our dataset (as described in
Section~\ref{sec:injection}) with varying ephemerides.  We vary the
injected transit ephemeris from 90 minutes before to 45 minutes after
the expected time of  transit, and we observe how the extracted
transmission spectrum varies.  These variations demonstrate the extent
to which temporal evolution of systematic effects may limit our
sensitivity. 

We find that the $\sigma_{d_\lambda}$ computed through this second
technique are generally larger than the uncertainties from our
residual-permutation analysis. To be conservative we assign the
differential transit depth in each wavelength channel an uncertainty
equal to the larger of the uncertainties from the two techniques.

\subsection{Comparing models to data}
\label{sec:comparing} We compare $d_\lambda$, the extracted transmission
spectrum, to the spectral transmission models described in
Section~\ref{sec:models} in two ways.  First, we cross-correlate our
extracted spectrum with each model and measure the integrated signal
centered at zero lag.  Second, we fit our extracted spectrum to (a)
each model spectrum and (b) a flat, featureless spectrum (representing
a nondetection) and compare the BIC of each fit.

We require that a robust detection of a given model must satisfy
several criteria.  First, we require that $d_\lambda$ is better fit
(in a least-squares sense) by the tested model than by a flat
spectrum.  Second, we require that the integrated cross-correlation
signal (i) is consistent with what we see when injecting synthetic
models into our data and (ii) is significantly greater than zero.  If
a particular model meets these criteria with the injected signals but
not with the unadulterated, observed data then we rule out the model
as a plausible description of \gjb 's atmosphere.  A model is
unconstrained if we are unable to reliably recover it when it has been
injected into the data.  Since we ultimately achieve no positive
detections, our main result is our ability to rule out a large segment
of atmospheric parameter space.

We do not use the entire set of extracted $d_\lambda$ when comparing
to models.  We exclude wavelength channels with S/N\,$< 40$, which
mainly affects spectral regions of low telluric or instrumental
transmission.  We also exclude all data from the first two echelle
orders ($\lambda < 2.1\micron$), because these wavelengths are heavily
corrupted by numerous, strongly saturated absorption features from
telluric \coo .  Finally, we exclude the region $2.196 \micron <
\lambda < 2.2 \micron$ because those portions of the data are visibly
corrupted by detector anomalies.  Because of the various models we
inject we do not end up using the same set of wavelength channels in
the analysis of each real or synthetic dataset, but in general we use
$4200 \pm 5$ (out of 7,168 possible) wavelength channels in the tests
that follow.

\subsubsection{Template cross-correlation}
\label{sec:xcorr} We first compare our extracted transmission spectrum,
$d_\lambda$, against atmospheric models by cross-correlating with the
model spectra described in Section~\ref{sec:models}.  We investigated
the use of the Least-Squares Deconvolution algorithm
\citep{donati:1997, barnes:2007a} to compare our measured transmission
spectrum with model spectra, but we find this method to be less
sensitive than straighforward cross-correlation.

Instead, we use the Discrete Correlation Function \citep[DCF;
][]{edelson:1988} to perform a cross-correlation between $d_\lambda$
and each spectral model.  The DCF has several advantages over standard
(linear-shift) methods for computing cross-correlations: it obviates
the need for interpolation to a linear wavelength grid, and it allows
us to compute the cross-correlation while excluding low S/N channels
without the need for interpolating across the resulting gaps.  In the
case of regularly sampled data the DCF gives results  identical to
linear-shift cross-correlation.

For each model transmission spectrum, we cross-correlate our extracted
transmission spectrum with the model over each echelle order.  We then
take the weighted average of these several cross-correlations, where
the weights are proportional to the number of wavelength channels used
to compute each echelle order's DCF.  

In the limit of high S/N and perfect correspondence between our model
and the observed transmission spectrum this analysis is equivalent to
computing the transmission spectrum autocorrelation.  We show the
autocorrelation for each echelle order, and the weighted average, in
Figure~\ref{fig:model_autocorr} for the case of our solar abundance
model.  The mean autocorrelation shows features on two characteristic
scales: a strong, narrow (tens of \kms) peak and a much broader
(roughly 200~\kms) plateau of lower amplitude; we refer to these as
the narrowband and broadband cross-correlation signals, respectively.
The former results mainly from the individual, narrow spectral
features while the latter results from broader variations of transit
depth with wavelength.  Integrating over these two correlation lag
scales provides us with two measures of the cross-correlation signal,
which we denote $\sum$DCF.  By varying the range of summation we
determined that we achieve the highest S/N with summation limits set
at $|v_{lag}| \leq 10$~\kms\ and $|v_{lag}| \leq 200$~\kms.

\fig{model_autocorr}{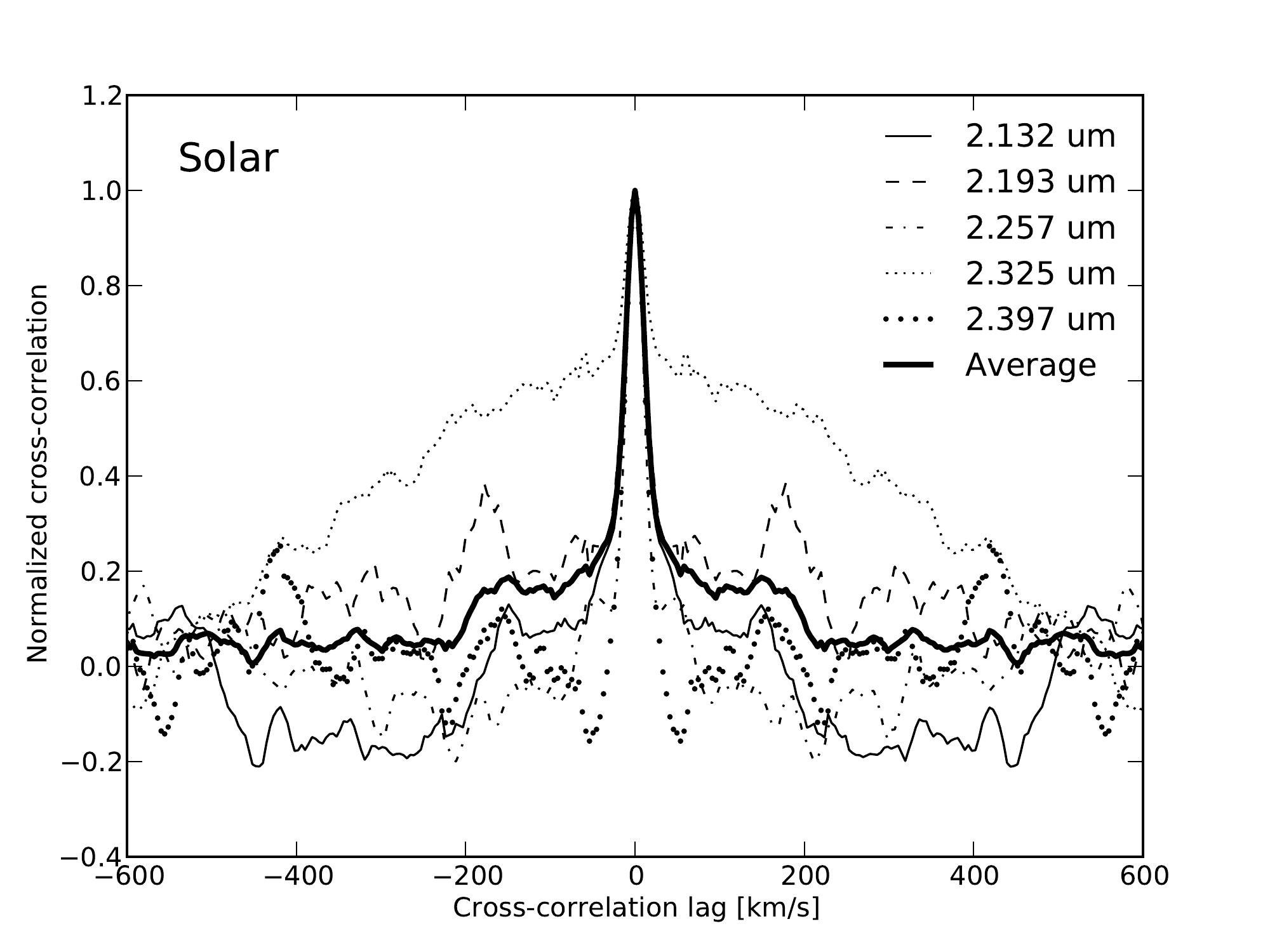}{width=7in}{}{Autocorrelation
  of our solar abundance model computed in each echelle order (thin
  lines), and the average (thick solid line) weighted by the number of
  wavelength channels used in our analysis.  All orders show a strong,
  narrow peak resulting from the narrow features in the model; broader
  structures (e.g. in the 2.325~\micron\ order) result from broad
  spectral variations (cf. Figure~\ref{fig:modelspec_zoom}).  In the
  limit of infinite S/N, this figure shows our expected
  cross-correlation signal.}

Examples of this cross-correlation process are shown in
Figures~\ref{fig:xcorr_solar}-\ref{fig:xcorr_last} for several of our
models.  These cross-correlations do not closely resemble the
noiseless autocorrelation shown in Figure~\ref{fig:model_autocorr}
because of residual instrumental and telluric sources of measurement
error.  We verified that, on average, the expectation value for the
effect of our noise sources on the cross-correlation values is zero
and thus the expectation values for our integrated signals are
unaffected.

\fig{xcorr_solar}{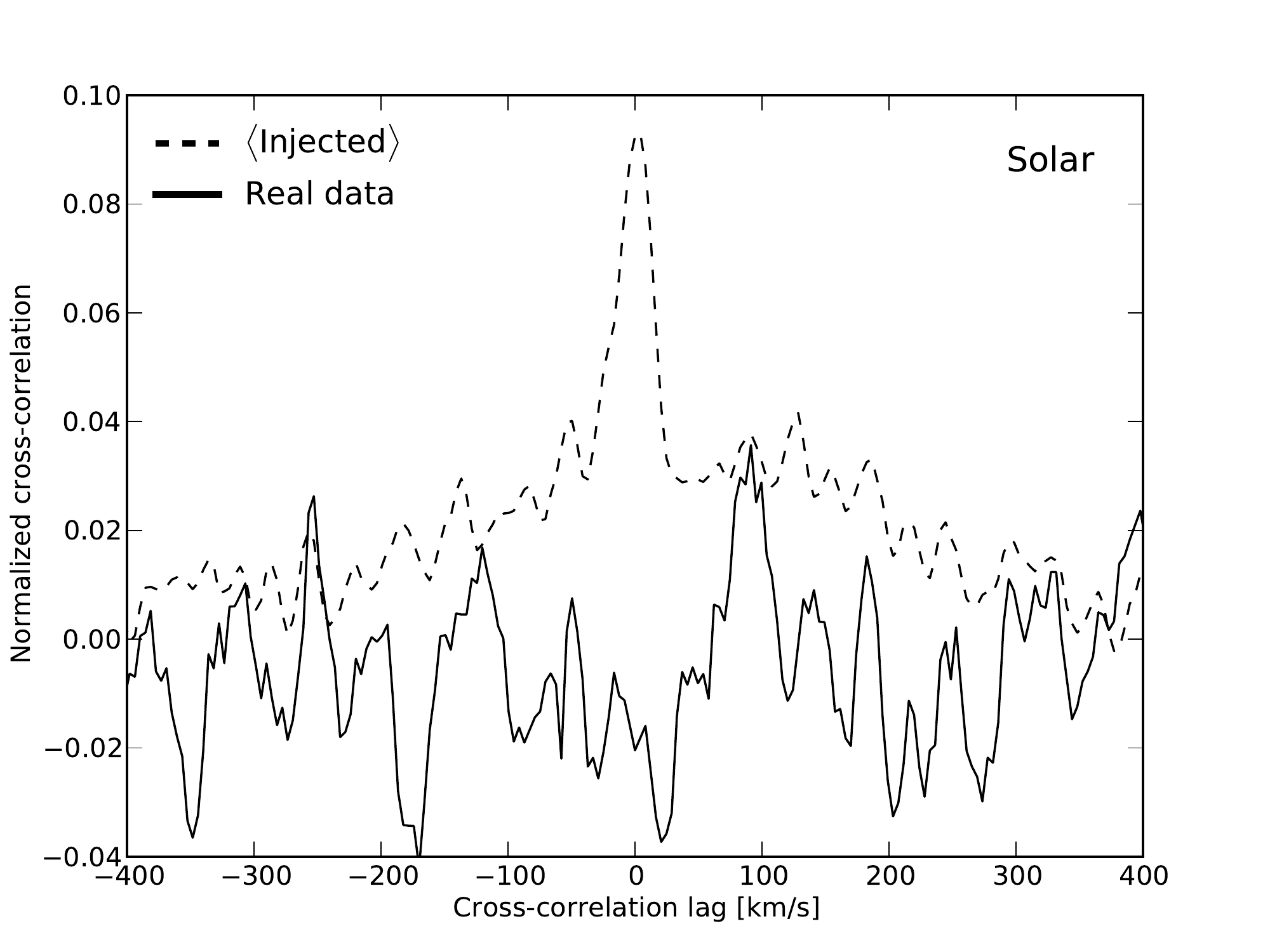}{width=7in}{}{Template
  cross-correlation results.  The solid line is the Discrete
  Correlation Function (DCF) of our solar-composition model with our
  extracted, unbinned transmission spectrum. The dashed line is the mean
  extracted DCF of the same model with data into which the transit
  signature of the model has been injected
  (cf. Section~\ref{sec:injection}). The integrated signal over our
  narrowband ($|v_{lag}| \leq 10$~\kms) and broadband ($|v_{lag}| \leq
  200$~\kms) apertures differ from the mean injected signal by
  4.6$\sigma$ and 5.6$\sigma$, respectively.  This allows us to rule
  out this model as a good representation of the transmission spectrum
  of \gjb.}

\fig{xcorr_10xsolar}{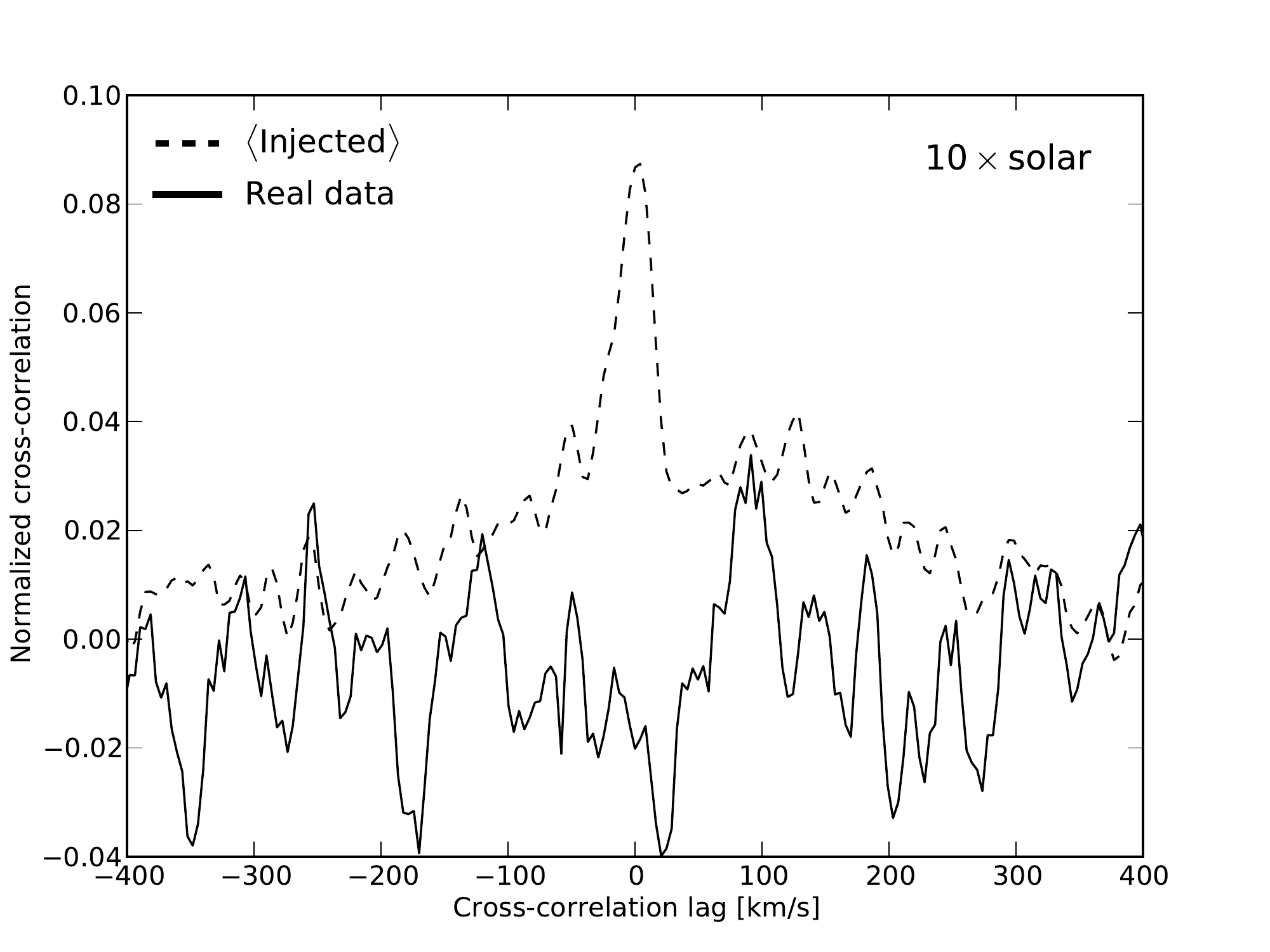}{width=7in}{}{Same as
  Figure~\ref{fig:xcorr_solar}, but for our 10$\times$ solar abundance
  model.  The narrowband and broadband integrated signals differ from
  the mean injected signal by 4.4$\sigma$ and 5.4$\sigma$,
  respectively.  This allows us to rule out this model as a good
  representation of the transmission spectrum of \gjb.}

\fig{xcorr_30xsolar}{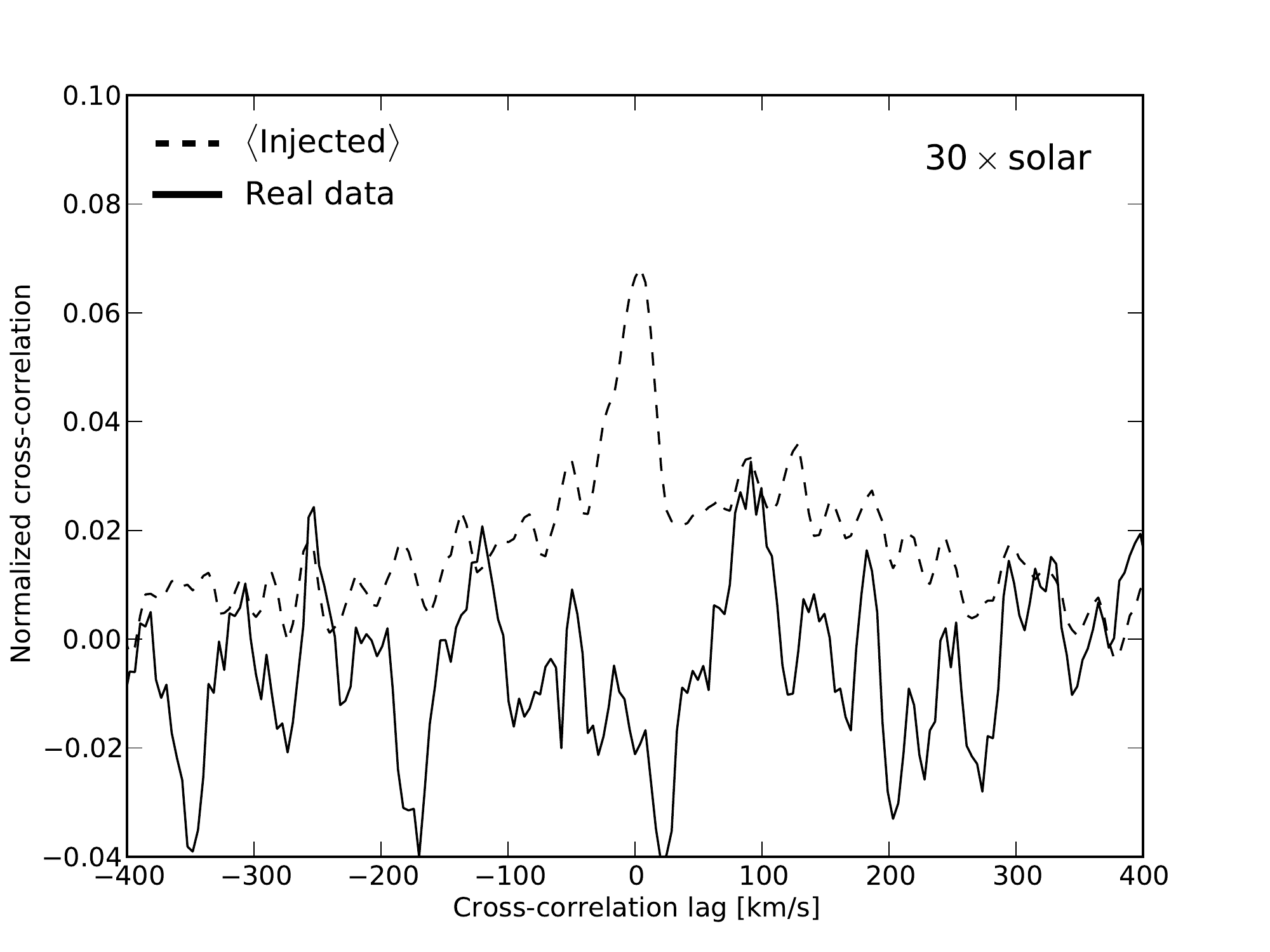}{width=7in}{}{Same as
  Figure~\ref{fig:xcorr_solar}, but for our 30$\times$ solar abundance
  model.  The narrowband and broadband integrated signals differ from
  the mean injected signal by 3.6$\sigma$ and 4.7$\sigma$,
  respectively.  This allows us to rule out this model as a good
  representation of the transmission spectrum of \gjb.}

\fig{xcorr_noch4}{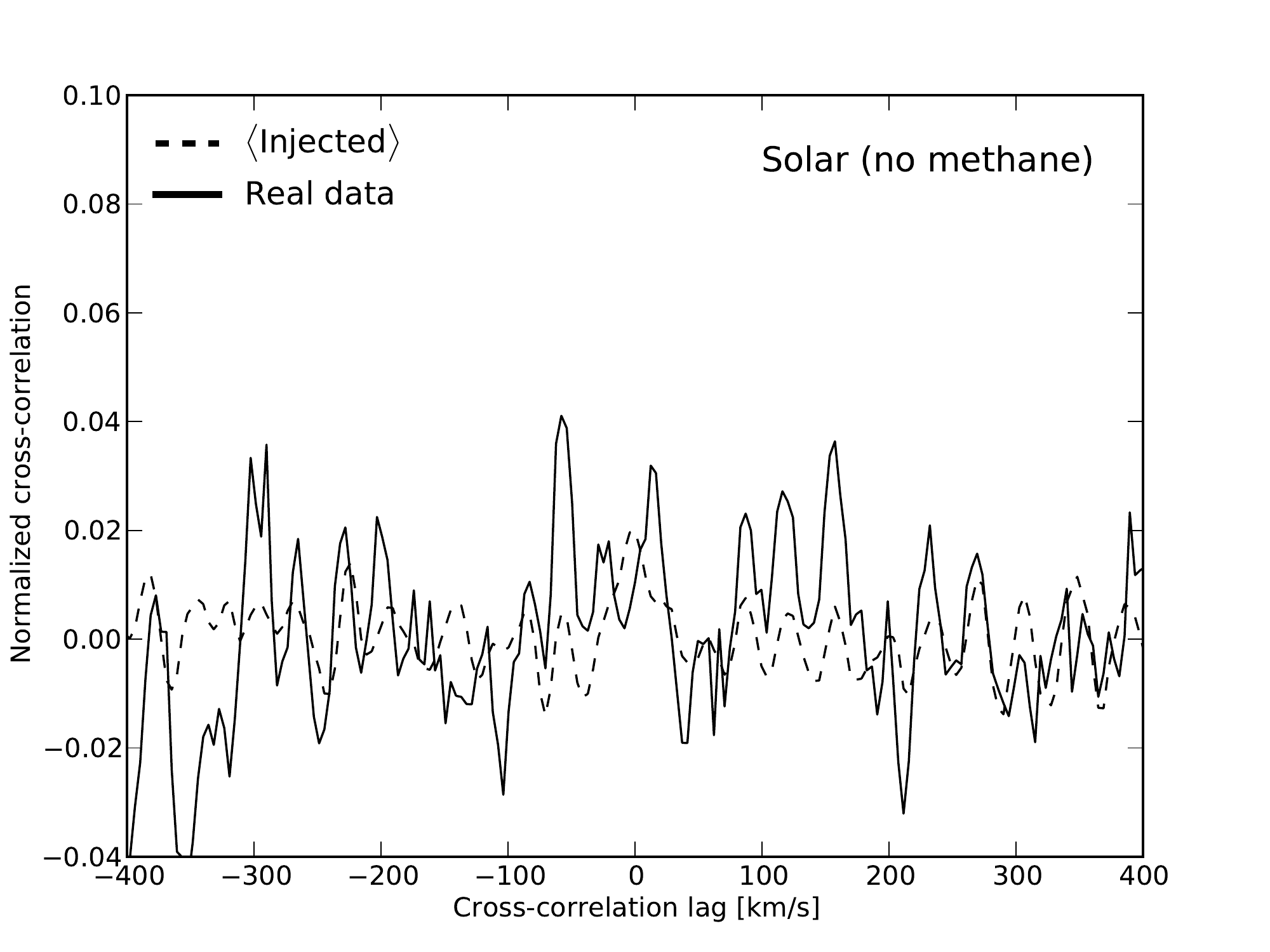}{width=7in}{}{Same as
  Figure~\ref{fig:xcorr_solar}, but for our solar abundance,
  methane-free model.  This model is quite flat and so we barely see a
  cross-correlation signal in the average, injected case; therefore
  our sensitivity does not allow us to  state with confidence whether such
  a model represents the transmission spectrum of \gjb.}

\fig{xcorr_last}{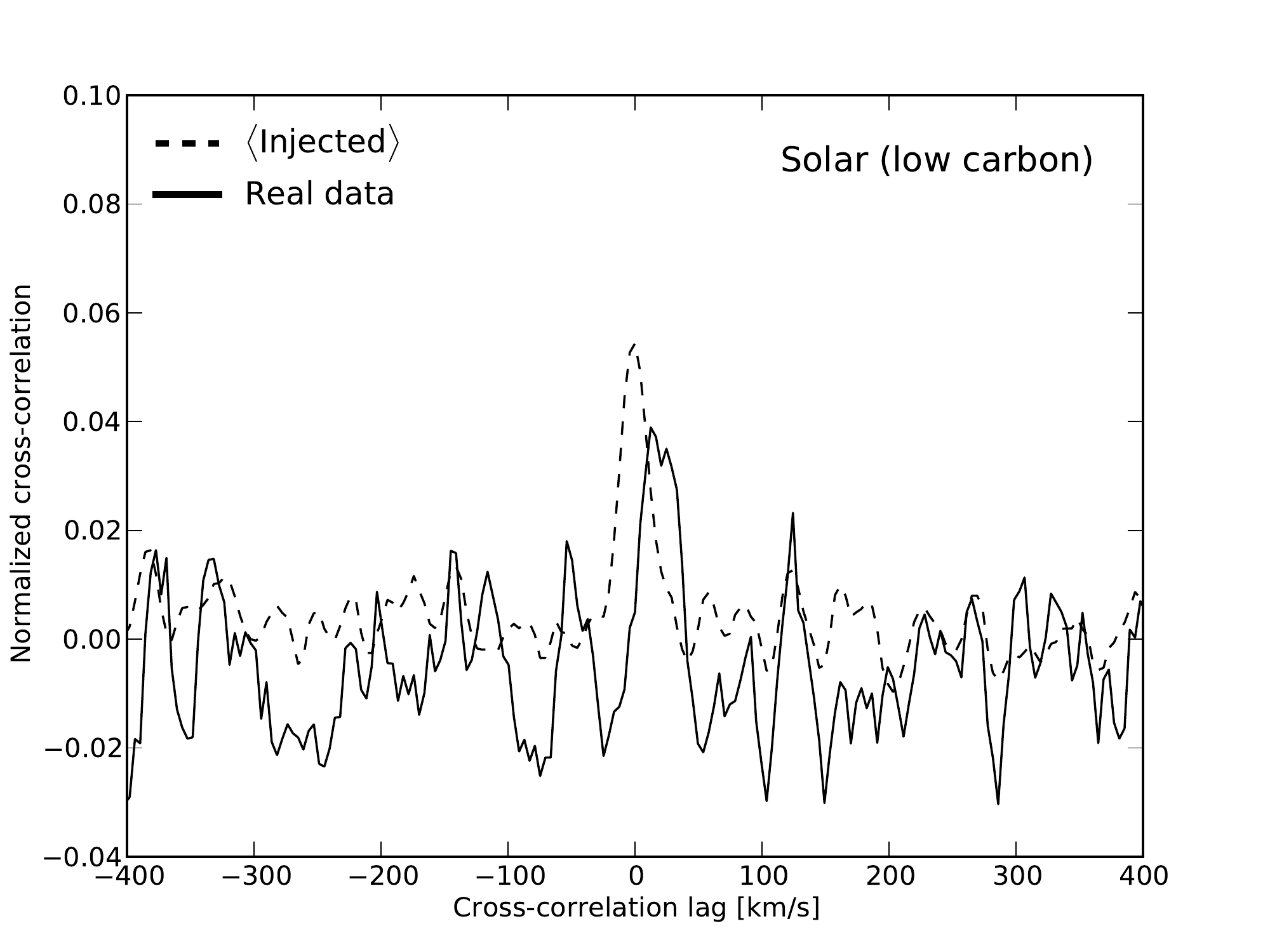}{width=7in}{}{Same as
  Figure~\ref{fig:xcorr_solar}, but for our low carbon abundance solar
  model.  The intriguing signal at $v\approx 20$~\kms\ has approximately
  the right shape and size as the expected signal but represents only
  a 1.9$\sigma$ detection; in addition its offset from zero lag seems
  too large to result from uncertainties in either \gjb 's orbit or in
  our wavelength calibration.  We deem this feature a false positive,
  and otherwise our sensitivity does not allow us to state with
  confidence whether such a model represents the transmission spectrum
  of \gjb.}

We quantify the significance of a $\sum$DCF measurement as follows.
We set our measurement uncertainty by taking the standard deviation of
the $\sum$DCF values measured for the case of an injected signal at
each of the several transit ephemerides (from 90~minutes earlier than,
to 45 minutes later than, the true time of transit).  We also
simulated cross-correlation functions by resampling a smoothed fit to
the mean power spectrum of the measured cross-correlation functions,
and then examining the frequency with which peaks of a particular
amplitude occur in this simulated data; these two techniques give
comparable results, which give us confidence in our measurement
uncertainty.  We then estimate the expected $\sum\textrm{DCF}$ value
in the presence of a known signal by injecting and re-extracting
models into our data at varying ephemerides (as described in
Section~\ref{sec:injection}).  We take the mean and standard error on
the mean of the $\sum\textrm{DCF}_\textrm{inj}$ to be our best
estimate of this quantity. We list all these values in
Table~\ref{tab:results}.  The agreement between the injected
($\sum$DCF$_\textrm{inj}$) and nominal ($\sum$DCF$_\textrm{0}$) values
indicates the degree to which the model corresponds to our
measurements; if they disagree at $>3\sigma$ we rule out the model
being tested.  Detection of a model would require the two
cross-correlation sums to be consistent and $\sum\textrm{DCF}_0$ to be
significantly greater than zero.

\subsubsection{Least-squares template matching}
We also compare the extracted spectrum, $d_\lambda$, to the various
spectral models using weighted linear least squares.  We determine the
offset value $C$ necessary to give the best match between the measured
and model spectra using $1/{\sigma^2_{d_\lambda}}$ as the weights --
that is, we solve $d_\lambda = m_\lambda - C$ in a least squares
sense, where $m_\lambda$ is the model spectrum being tested. The
offset $C$ results from our removal of the absolute transit depth by
our correction for slit losses. We compute the statistic
$\Delta\textrm{BIC}$ equal to the difference between the BIC for each
model and the BIC resulting from a flat, featureless spectrum (i.e., a
nondetection). Then, $\Delta\textrm{BIC} \le 0$ implies the data are
more consistent with a constant $R_P(\lambda)$ than with the tested
atmospheric model, while $\Delta\textrm{BIC} > 0$ indicates we are
justified in preferring the additional parameter -- i.e., the model
spectrum -- to match our data.  We list the $\Delta$BIC from comparing
each of our spectral models in Table~\ref{tab:results}.

\subsection{Sensitivity tests: injected models}
\label{sec:injection} 
As previously alluded to, we test the sensitivity of our analysis by
inserting our model transmission spectra (cf.
Section~\ref{sec:models}) into our extracted echelle spectra and
attempting to recover the injected signal.  We include both the
time-dependent wavelength shift of the transmission spectrum due to
the planet's orbital motion and the overall transit light curve shape
in each wavelength channel.  Our ability to recover these
velocity-shifted models justifies our decision to neglect the radial
velocity shift when fitting each wavelength transit events.  All data
outside of the simulated transit are unaffected. 

We inject these signals using ephemerides shifted by up to 90~minutes
earlier than, and 45 minutes later than, the true transit ephemeris at
intervals of 15 minutes.  We inject each model spectrum immediately
before the echelle spectra have been brought to a common wavelength
frame (i.e., at the beginning of Section~\ref{sec:sysrem}) and repeat
the same analyses and tests as described in the preceding sections.
The mean and error on the mean of the ensemble of the results of these
injected and re-extracted spectra are listed in
Table~\ref{tab:results}; these define our sensitivity limits.

\begin{deluxetable}{l c | c c | r r | r r | l} 
\tablewidth{0pt}
\tabletypesize{\footnotesize}
  \tablecolumns{8} \tablecaption{Results from spectral model testing \label{tab:results}}
\tablehead{  & & \multicolumn{2}{c  }{} & \multicolumn{2}{c  }{$|v_\textrm{lag}|\le 10$~\kms} & \multicolumn{2}{ c  }{$|v_\textrm{lag}|\le 200$~\kms} \\ 
\colhead{Model}  & \colhead{$\sigma_{R_P}$\tablenotemark{a}} &  \colhead{$\Delta$BIC$_0$\tablenotemark{b}} & \colhead{$\langle\Delta$BIC$_\textrm{inj}$\tablenotemark{b}$\rangle$} & 
\colhead{$\sum_{v}$DCF$_0$\tablenotemark{c}}   &  \colhead{$\langle\sum_v$DCF$_\textrm{inj}\rangle$\tablenotemark{d}}  &
\colhead{$\sum_{v}$DCF$_0$\tablenotemark{c}}   &  \colhead{$\langle\sum_v$DCF$_\textrm{inj}\rangle$\tablenotemark{d}}  &
  \colhead{Result\tablenotemark{e}}}
\startdata

                 30$\times$ solar  & 0.058 & -180 & 177 & -0.35(43) &  1.33(17)  &  -2.0(2.4) &  10.2(1.0) & Ruled out\\
        ''   ($\log_{10}({\textrm{\methane}})\approx -3.0 $) & 0.052 & -150  & 148 & -0.35(43) &  1.20(16)  &  -2.0(2.4) &   9.4(1.0)   & Ruled out\\
        ''  ($\log_{10}({\textrm{\methane}})\approx -4.0 $) & 0.047 & -124  & 122 &  -0.34(43) &  1.07(16)  &  -1.9(2.5) &   8.7(1.0)  & Ruled out\\
        ''  ($\log_{10}({\textrm{\methane}})\approx -5.1 $) & 0.041 & -101  & 99 &  -0.34(43) &  0.95(16)  &  -1.7(2.5) &   7.9(1.0) & Unconstrained \\
                           
\hline
                 10$\times$ solar  & 0.076  & -288  & 285 &  -0.33(43) &  1.70(17)  &  -2.1(2.5) &  12.5(1.1) & Ruled out \\
     ''  ($R_{min} = 2.85$~\rearth)  & 0.071  & -255  & 234 &  -0.37(44) &  1.58(18)  &  -2.5(2.7) &  11.8(1.2) & Ruled out\\
     '' ($R_{min} = 2.9$~\rearth) & 0.064   & -205  & 176 &  -0.44(47) &  1.39(19)  &  -3.1(2.8) &  10.7(1.2) & Ruled out\\
       ''  ($\log_{10}({\textrm{\methane}})\approx -3.6 $)  & 0.069  & -238  & 237 & -0.33(43) &  1.54(17)  &  -2.0(2.5) &  11.5(1.1) & Ruled out\\
       ''  ($\log_{10}({\textrm{\methane}})\approx -4.6 $)  & 0.061  & -194  & 194 & -0.32(43) &  1.37(17)  &  -1.9(2.6) &  10.5(1.1) & Ruled out\\
       ''  ($\log_{10}({\textrm{\methane}})\approx -5.6 $)  & 0.054  & -155  & 156 & -0.31(43) &  1.21(16)  &  -1.8(2.6) &   9.5(1.1) & Ruled out \\
       ''  ($\log_{10}({\textrm{\methane}})\approx -6.6 $)  & 0.047  & -122  & 122 & -0.30(43) &  1.05(16)  &  -1.6(2.7) &   8.4(1.1) & Unconstrained \\

\hline

              Solar    & 0.082  & -329  & 315  & -0.34(43) &  1.81(18)  &  -2.4(2.5) &  13.1(1.1) & Ruled out\\ 
   ''   ($R_{min} = 2.85$~\rearth) & 0.081  & -320   & 300   & -0.34(43) &  1.78(18)  &  -2.5(2.6) &  12.9(1.1) & Ruled out\\
   ''   ($R_{min} = 2.9$~\rearth)    & 0.075  & -291  & 193 & -0.39(45) &  1.46(19)  &  -2.9(2.7) &  10.8(1.4) & Ruled out\\
   
   ''     ($\log_{10}({\textrm{\methane}})\approx -3.8 $) & 0.074  & -273  & 263 & -0.33(43) &  1.63(18)  &  -2.4(2.6) &  12.1(1.1)  & Ruled out\\
   ''     ($\log_{10}({\textrm{\methane}})\approx -4.7 $) & 0.067  & -224  & 216 & -0.33(43) &  1.45(17)  &  -2.3(2.6) &  11.0(1.1)  & Ruled out\\
   ''     ($\log_{10}({\textrm{\methane}})\approx -5.5 $) & 0.059  & -179  & 174 & -0.32(43) &  1.28(17)  &  -2.3(2.7) &   9.9(1.1)  & Ruled out\\
   ''     ($\log_{10}({\textrm{\methane}})\approx -6.4 $) & 0.052  & -141  & 137 & -0.31(43) &  1.10(17)  &  -2.2(2.7) &   8.8(1.1)  & Ruled out\\
   ''     ($\log_{10}({\textrm{\methane}})\approx -7.3 $) & 0.045  & -107  & 104 & -0.30(43) &  0.93(16)  &  -2.0(2.8) &   7.7(1.1)  & Unconstrained\\

    '' (low carbon)  & 0.050  & -61  & 36 &  0.20(34) &  0.99(13)  &  -1.5(1.7) &   2.6(0.6) & Unconstrained \\
    '' (low C; $v_0=20$~\kms)\tablenotemark{f}  & 0.050  & -61  & 36 &  0.72(38) &  0.99(13)  &  -1.6(1.6) &   2.6(0.6) & Unconstrained \\
    '' (no methane)  & 0.026  & -14  & 6.1 &  0.22(49) &  0.35(17)  &   2.1(3.1) &   0.1(0.1)   & Unconstrained \\

                              
\enddata                       

\tablenotetext{a}{Standard deviation of the model $R_P(\lambda)$, in
  Earth radii, over the wavelengths used in our analysis and at our
  model resolution.}

\tablenotetext{b}{$\Delta\textrm{BIC} = \textrm{BIC}_\textrm{flat} -
  \textrm{BIC}_\textrm{model}$; a positive value implies that the the
  extracted transmission spectrum is better fit by the given spectral
  model than by a flat, featureless spectrum;
  cf. Section~\ref{sec:detection}.}

\tablenotetext{c}{Cross-correlation sum for the nominal case -- i.e.,
  the true transit ephemeris and without any injected signal.  The
  uncertainty quoted is the standard deviation of the quantities
  measured at alternate transit ephemerides.}

\tablenotetext{d}{Expectation value in the case of a positive signal,
  estimated by calculating the mean and standard deviation on the mean
  of the values measured for the injected cases.}

\tablenotetext{e}{To rule out a model we require that $\Delta$BIC$>0$
  in the injected case and that $\sum_v\textrm{DCF}_0 < \langle
  \sum_v\textrm{DCF}_\textrm{inj} \rangle$ at greater than $3\sigma$ confidence
  for both the narrowband and broadband cross-correlation signals.}

\tablenotetext{f}{{Low} carbon model, but with the limits of
  cross-correlation integration shifted by 20~\kms\ to encompass the
  feature centered there (cf. Figure~\ref{fig:xcorr_last}).  The model
  is still unconstrained in this case because the narrowband
  $\sum$DCF$_0$ is still $<2\sigma$ discrepant from zero. }

\end{deluxetable}

\section{Results and  Discussion}
\label{sec:results}
\subsection{Spectroscopic results}
We see no evidence of a significant match between our extracted
transmission spectrum and any of our various spectral models. These
nondetections are significant because our sensitivity analysis
described above demonstrates our ability to recover signals of
comparable magnitude in our data, as demonstrated by the values of
$\Delta$BIC and $\sum$DCF listed in Table~\ref{tab:results} for each
model.  In this section we discuss the results of our spectroscopic
analysis, and in the section that follows we discuss these results in
the context of other observations.

We rule out cloud-free atmospheres in chemical equilibrium with solar,
10$\times$ solar, and 30$\times$ solar abundances.  Additionally, we
rule out clear atmospheres with \methane\ abundances reduced from that
expected in chemical equilibrium: we constrain
$\log_{10}(N_\textrm{CH4})$, the \methane\ molar abundance, to be
$<$~$-6.4$, $<$~$5.6$, and $<$~$4.0$ in the case of solar, 10$\times$
solar, and 30$\times$ solar abundance, respectively.  In general we
have less ability to discriminate between variants of the 30$\times$
solar model than of the solar abundance model.  This is because the
variations of transit depth with wavelength decrease with increasing
atmospheric metallicity, as seen in Figure~\ref{fig:modelspec_zoom},
because a greater mean molecular weight results in a smaller scale
height and thus a general suppression of the amplitudes of all
spectral features. A similar trend is also apparent in the models of
\cite{miller-ricci:2010} with atmospheres even more enriched in heavy
elements. We demonstrate an ability to constrain the presence of
clouds: in the case of an atmosphere with solar or 10$\times$ solar
abundance, an opaque cloud deck as high as $2.9$~\rearth (roughly
100~mbar; cf. Figure~\ref{fig:atmo}) still allows us to detect the
spectroscopic signature of the planet.

The closest we come to detecting any atmospheric model (i.e., those
tests which result in the largest values of $\Delta$BIC and
$\sum$DCF$_0$) are our no-methane and low-carbon models.  The
no-methane model (cf. Figure~\ref{fig:xcorr_noch4}) is the only model
with a broadband ($|v_{lag} \leq 200$~\kms) sum greater than zero, but
it is still consistent with zero at the 0.7$\sigma$ level.
Cross-correlation with the low-carbon model
(cf. Figure~\ref{fig:xcorr_last}) produces a very interesting feature,
similar in size and shape to the expected narrowband signal but offset
by 20~\kms. However, this large cross-correlation lag seems too large
to be explained by either uncertainties in \gjb 's orbit or in our
wavelength calibration.  Since this feature is only discrepant from
zero at the 1.9$\sigma$ level, both these low-significance $\sum$DCF
measurements seem most likely to be simple false positives.
Nonetheless, these low-methane models remain intriguing in light of
the correspondence between their predicted planetary radii and recent
photometric observations (as we describe below).

\subsection{Constraints on \gjb 's atmosphere}
If \gjb\ has a H$_2$/He-dominated atmosphere, our observations require
a reduced methane abundance in order for the atmospheric signature to
remain undetectable to our analysis.  Such an atmosphere also resolves
the tension between the recent claims of a high
\citep{bean:2010,desert:2011} and low \citep{croll:2011b} mean
molecular weight atmosphere on \gjb .  \cite{zahnle:2009} has
suggested that the atmospheres of cooler (but still strongly
irradiated) planets may be conducive to photochemistry and
polymerization of \methane\ into more complex organic compounds, which
may form a haze with significant optical opacity.  An optical haze
\citep[as observed on HD~189733b;][]{sing:2011} formed from \methane\
reaction products provides a possible way to reconcile the nearly flat
optical \citep{bean:2010} transmission spectrum with the significantly
larger Ks-band radius \citep{croll:2011b}.

Thus a methane-depleted, H$_2$/He-dominated atmosphere with an optical
haze appears broadly consistent with all of the optical and infrared
results for this system.  Intriguingly, the only two atmospheric
models to evince even a hint of a detected signal in our spectroscopic
analysis have little or no methane; these are the models with low
carbon and zero methane abundances (cf. Table~\ref{tab:results}, and
Figures~\ref{fig:xcorr_noch4} and~\ref{fig:xcorr_last}); the
correspondence is curious and warrants further consideration. As
Figure~\ref{fig:ratios} demonstrates, both of these models are
consistent (within 1.5$\sigma$) with the observed mid- and
near-infrared planetary radius ratios \citep{desert:2011,croll:2011b}
-- compare this to the equilibrium-abundance models ruled out by our
analysis, which are also discrepant (at $>5\sigma$) from the
photometric observations. Further observations are certainly needed.
Fortunately, a methane-depleted atmosphere should be easy to
characterize since it should still exhibit prominent absorption
features from \water .  To disentangle the effects of these
exoplanetary absorption features from the telluric \water\ that
defines the edges of ground-based observing bands, the atmosphere
might best be characterized from space using the HST WFC3 grism in the
J and H bands; ground-based narrowband infrared transit photometry
(e.g., 2.1-2.3 \micron) could also test the existence of such an
atmosphere.

\fig{ratios}{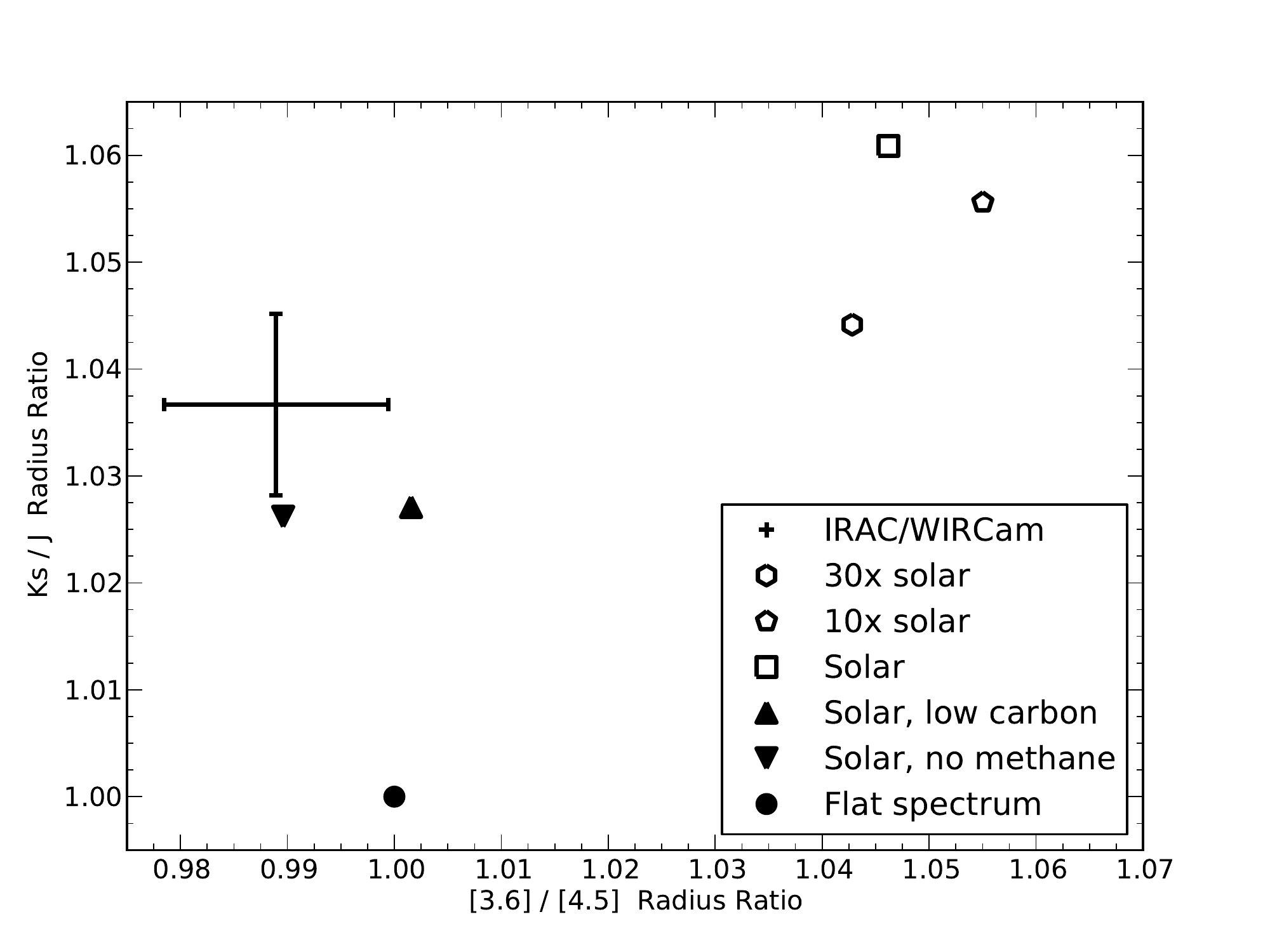}{width=7in}{}{Near-infrared (Ks/J) and
  mid-infrared ([3.6]/[4.5]) planetary radius ratios for \gjb . Values
  predicted by our atmospheric models, and the observed ratios and
  their uncertainties as measured with IRAC \citep{desert:2011} and
  WIRCam \citep{croll:2011b}, are plotted.  Solid symbols are models
  consistent with our analysis; open symbols are models that we rule
  out. Models subtantially depleted in methane are consistent with our
  and prior observations, but equilibrium-abundance H$_2$/He-dominated
  models are strongly inconsistent with all observations.  A flat
  transmission spectrum (due, e.g., to an atmosphere dominated by
  heavier molecular species), though consistent with our spectroscopic
  analysis, is inconsistent with photometric observations at
  $>4\sigma$. }

A number of possible interpretations consistent with our results and
with other observations are excluded solely by the large Ks transit
depth \citep{croll:2011b}.  For example, our data alone are consistent
with a dense molecular (non-H$_2$) atmosphere on \gjb, which would
reduce the atmospheric scale height such that the transmission
spectrum would be essentially constant with wavelength.  This is
consistent with the near-constant optical and mid-infrared transit
depth \citep{bean:2010,carter:2011,desert:2011}.  This scenario would
imply that \gjb\ probably did not accrete much of its atmosphere from
the gaseous protoplanetary disk in which it presumably formed, but
instead formed largely from accretion of, and outgassing by,
volatile-rich ices \citep{rogers:2010}.  Such an atmosphere would have
features in transmission an order of magnitude weaker than those
expected for a H$_2$/He-dominated atmosphere, and so would be more
amenable to characterization by looking for the $\lesssim10^{-3}$
mid-infrared ($>$5~\micron) secondary eclipse signature
\citep{miller-ricci:2010}.  Such observations may have to wait for the
James Webb Space Telescope, though perhaps dedicated observations with
the Stratospheric Observatory for Infrared Astronomy (SOFIA) could
make a contribution somewhat sooner.

\gjb 's atmosphere would also be consistent with our observations if
it were covered in high-altitude ($>2.9$~\rearth\,\ or $<100$~mbar)
clouds that would mask the spectroscopic signals to which we are
sensitive.  However, the effect of such a cloud layer would presumably
extend to the optical, where observations demonstrate that the planet
is significantly smaller than this.  This discrepancy suggests that in
a H$_2$/He-dominated atmosphere clouds could not consistently explain
our (and prior) results. 

In our calculations we have taken the larger stellar radius
(0.21\rsun) proposed by \cite{carter:2011}.  It is possible that this
overestimates the radius of the host star by as much as 15\%
\citep{charbonneau:2009,carter:2011}, making any planetary atmosphere
more difficult for us to detect.  A smaller radius would increase the
surface gravity and decrease the atmospheric scale height, and a
higher density would make the planet somewhat more consistent with
denser interior models that may be less likely to host the extended,
H$_2$/He-dominated envelopes to which our observations are most
sensitive.  This reduced radius would allow \gjb 's mass and radius to
be fit without any atmosphere whatsoever \citep[even the larger radius
is only 2$\sigma$ discrepant from such a
scenario;][]{rogers:2010}. Though it may be difficult to motivate a
formation and evolution scenario that leads to a composition wholly
lacking an atmosphere, this would be consistent with most current
observations -- except, once again, for the Ks band transit
measurements \citep{croll:2011b} which seems to argue against all
scenarios but those with H$_2$/He-dominated atmospheres.

\subsection{Lessons for future ground-based spectroscopy}
High-resolution spectroscopy holds promise for the characterization of
extrasolar atmospheres, as evidenced by the limits we set here with
only half a night's observations.  Throughout this paper, we have
noted that our analysis is sensitive to variations in transit depth
with wavelength, and not to the absolute transit depth.  A simple way
to quantify our final sensitivity is in terms of the level of radius
variability we can rule out, as measured by $\sigma_{R_P}$, the
standard deviation of $R_P(\lambda)$ at our model resolution. Though
the details of our sensitivity analysis depend on the shape of the
spectrum being tested, a simple rule of thumb turns out to work fairly
well: as seen from Table~\ref{tab:results}, we can reliably detect
models with $\sigma_{R_P} \gtrsim 0.05$~$\rearth$.  Equivalantly, any
tested model with wavelength-dependent transit depth variations
$\gtrsim 5 \times 10^{-4}$ over our wavelength range should be
detectable to our analysis.  Successful characterization with NIRSPEC
of a 100\% \water\ atmosphere on a \gjb-like planet would require roughly a
factor of 5 better precision, which seems at the outer edge of
what might be achieved with a dedicated spectroscopic campaign.  The
atmospheres of smaller, cooler, and more Earth-like planets will be
even more challenging to detect. On the other hand, a GJ~436b-like Hot
Neptune would have an atmospheric signal only a factor of two below
our sensitivity, and so such an object might be amenable to
high-resolution spectroscopic characterization in only a single
observing season.

Our sensitivity is fundamentally limited by the spectrophotometric
variability resulting from (a) variations in spectroscopic slit loss,
(b) telluric-induced flux variations, and (c) slow drifts likely of
instrumental or residual telluric origin.  We address the first by
removing the common-mode slitloss term (removing the absolute transit
depth from our data), the second by calibrating our data with various
empirical airmass terms, and the third by including low-order
polynomials in our fitting process; our ability to address these
suggests that further observations with NIRSPEC or comparably
instruments would probably allow us to probe a somewhat greater region
of atmospheric parameter space and should not be discounted (we are
aware of at least one other group which has obtained additional
NIRSPEC K band observations of \gjb\ during transit, and we look
forward to seeing whether these can more tightly constrain the model
parameter space we have explored).

That said, we note that concerns (a) and (b) above can be largely
eliminated -- without the penalties or tradeoffs incurred by our
methods -- by multi-object spectroscopy \citep{bean:2010}; this
technique would at least partially mitigate limitation (c) as well,
though there is some evidence that field-rotating multi-object
spectrographs may suffer their own peculiar set of systematic effects
\citep{moehler:2010}.  There are a growing number of cryogenic,
infrared multi-object spectrographs on large-aperture telescopes:
MOIRCS at Subaru, and soon MOSFIRE at Keck and FLAMINGOS-2 at Gemini
South.  Combined with extant optical multi-object spectrographs these
instruments may have a leading role to play in the future
characterization of exoplanetary atmospheres.

A subset of exoplanet systems could be amenable to spectroscopic
characterization in an alternate manner.  If an exoplanet host star
has another star of comparable brightness nearby, observations from a
sufficiently precise slit-viewing camera or facility guiding camera
could be used to calibrate out the (broadband) photometric variations
resulting from telluric effects.  With a sufficiently wide
spectrograph slit (probably $\ge3$''), good guiding, and no slit
nodding the slit loss term that limits our observations here could
also be almost wholly compensated for.  The independent photometric
calibration offered by this technique would also be useful in refining
the precise transit ephemeris of a given observation and of
characterizing any spot crossings or other astrophysical``red noise''
sources. This method could enable at least coarse spectroscopic
characterization on even smaller (and hence, more accessible)
telescopes, and we plan to test it in future observations.

\section{Conclusions}
\label{sec:conclusion} Planets with \gjb 's mass and radius have no solar
system analogues, so {\em a priori} we know relatively little about
the nature of their interiors or atmospheres.  Here we report
high-resolution near-infrared spectroscopy of the \gj\ system during
\gjb 's transit, taken to ascertain the nature of the planet's
atmosphere.  Calibration using a common-mode flux variation time
series (which removes the absolute transit depth, $C$, from the data)
allows us to search for the relative change in transit depth with
wavelength: that is, we measure the quantity $(R_P(\lambda)/R_*)^2 -
C$.

Our spectroscopy rules out a clear, cloudless atmosphere in or near
chemical equilibrium for \gjb\ assuming solar, 10$\times$ solar, and
30$\times$ solar abundances, and in this sense we are consistent with
previous results \citep{bean:2010,desert:2011}.  Our data are
consistent with any of the following: (i) a H$_2$/He-dominated
atmosphere depleted in methane, (ii) an atmosphere dominated by
heavier molecular species with a smaller atmospheric scale height, or
(iii) almost any atmospheric composition if obscured by a
high-altitude cloud layer.  However, the large Ks band transit depth
reported for this system \citep{croll:2011b} argues against (ii) and
(iii) above.  \gjb\ thus appears more likely to be a scaled-down
version of GJ~436b (another methane-depleted, H-dominated planet
orbiting an M dwarf) than a rock- or ice-dominated body with a dense
envelope.  As the interpretation hinges largely on the near- and
mid-infrared measurements, further observations of this system are
needed to pin down the nature of its atmosphere.

Planet surveys suggest that objects of this mass and size may be quite
common \citep{howard:2010, borucki:2011}, so we expect similar planets
to be confirmed in the future.  Further observations of \gjb\ during
transit -- both photometric and spectroscopic -- will help refine our
understanding of this object and future objects like it, and pave the
way toward the characterization of ever more Earthlike planets.
Because \gj\ is such a cool star, infrared wavelengths continue to
offer the best prospects for these future observations.  We suggest
that infrared spectroscopy (especially from spectrographs such as
Subaru/MOIRCS, Keck/MOSFIRE, and HST/WFC3) and narrowband photometry
will be able place the tightest atmospheric constraints on this planet
and those like it as yet undiscovered.

\section{Acknowledgements}
Support for this work at UCLA was provided by NASA through awards
issued by JPL/Caltech and the Space Telescope Science Center.  TB is
supported by a NASA Origin of Solar Systems grant awarded to Lowell
Observatory.  We thank I. McLean for useful discussions about the
NIRSPEC instrument, D. Rodriguez for helpful comments during
manuscript preparation, B. Croll for useful feedback on the
manuscript, and our anonymous referee for a through and detailed
report that improved this paper.

The data presented herein were obtained at the W.M. Keck Observatory,
which is operated as a scientific partnership among the California
Institute of Technology, the University of California and the National
Aeronautics and Space Administration. The Observatory was made
possible by the generous financial support of the W.M. Keck
Foundation.  The authors wish to recognize and acknowledge the very
significant cultural role and reverence that the summit of Mauna Kea
has always had within the indigenous Hawaiian community.  We are most
fortunate to have the opportunity to conduct observations from this
mountain.

\footnotesize

\bibliographystyle{apj_hyperref}
\bibliography{ms}



\end{document}